\documentclass[fleqn,a4paper]{article}

\usepackage{a4}
  \usepackage{amsmath}
  \usepackage{amsthm}
  \usepackage{amssymb}
  \usepackage{graphicx}
  \usepackage{stmaryrd}
  \usepackage{color}
  \usepackage{graphs}
  \newcommand{\IFF}{iff}

  \newcommand{\is}[1]{\ensuremath{\mathalpha{\exists{#1}}}}

  \newcommand{\pequate}{\ensuremath{\mathrel{\simeq}}}
  \renewcommand{\merge}{\mathrel{\parallel}}
  \newcommand{\bisim}[1][]{%
    \setbox0=\hbox{\kern-.1ex{$\leftrightarrow$}\kern-.1ex}
    \setbox1=\vbox{\hbox{\raise .1ex \box0}\hrule}%
    \ensuremath{\mathrel{\hbox{\kern.1ex\box1\kern.1ex}_{#1}}}
  }
  \newcommand{\bbisim}{\bisim_{\text{b}}}
  \newcommand{\wbisim}{\bisim_{\text{w}}}
  \newcommand{\Act}{\ensuremath{\mathcal{A}}}
  \newcommand{\Actt}{\ensuremath{\Act_\tau}}
  \newcommand{\nil}{\ensuremath{\mathalpha{\mathbf{0}}}}
  \newcommand{\pref}[1]{\ensuremath{\mathalpha{{#1}.}}}
  \newcommand{\starpref}[1]{\ensuremath{\mathalpha{{#1}^{*}}}}
  \newcommand{\altc}{\mathbin{+}}
  \newcommand{\eqclass}[1]{\ensuremath{[#1]}}
  \newcommand{\beqclass}[1]{\ensuremath{[#1]_{b}}}
  \newcommand{\weqclass}[1]{\ensuremath{[#1]_{w}}}
  \newcommand{\algel}[1]{\ensuremath{\mathalpha{#1}}}
  \newcommand{\facalg}[2]{\ensuremath{{#1}{/}{#2}}}

  \newcommand{\act}[1][\alpha]{\ensuremath{\mathalpha{#1}}}
   \newcommand{\acta}[1][]{\act[a_{#1}]}
   \newcommand{\actb}[1][]{\act[b_{#1}]}
   \newcommand{\silent}{\act[\tau]}
  \newcommand{\PTERMS}[1][P]{\ensuremath{\mathcal{#1}}}
  \newcommand{\pterm}[1][P]{\ensuremath{\mathalpha{#1}}}
   \newcommand{\ptermP}[1][]{\pterm[P_{#1}]}
   \newcommand{\ptermQ}[1][]{\pterm[Q_{#1}]}
   
  \newcommand{\CPTERMS}[1][P]{\ensuremath{\mathcal{#1}}}
  \newcommand{\cpterm}[1][P]{\ensuremath{\mathalpha{#1}}}
   \newcommand{\cptermP}[1][]{\cpterm[P_{#1}]}
   \newcommand{\cptermQ}[1][]{\cpterm[Q_{#1}]}
   \newcommand{\cptermR}[1][]{\cpterm[R_{#1}]}
  
  \newcommand{\BPROC}[1][B]{\ensuremath{\mathalpha{\mathbf{#1}}}}
  \newcommand{\BPROCtn}[1][B]{\ensuremath{\mathalpha{\mathbf{#1}_{\textit{n}}}}}
  \newcommand{\BPROCfin}[1][B]{\ensuremath{\mathalpha{\mathbf{#1}_{\textit{fin}}}}}
  \newcommand{\WPROC}[1][W]{\ensuremath{\mathalpha{\mathbf{#1}}}}
  \newcommand{\WPROCtn}[1][W]{\ensuremath{\mathalpha{\mathbf{#1}_{\textit{n}}}}}
  \newcommand{\WPROCfin}[1][W]{\ensuremath{\mathalpha{\mathbf{#1}_{\textit{fin}}}}}

  \newcommand{\proc}[1][p]{\ensuremath{\mathalpha{#1}}}
   \newcommand{\procp}[1][]{\proc[p_{#1}]}

  \newcommand{\brelsym}{\ensuremath{\mathalpha{\mathcal{R}}}}
  \newcommand{\brel}{\ensuremath{\mathrel{\brelsym}}}

  \newcommand{\decompsym}{\ensuremath{\partial}}
  \newcommand{\decomp}[1]{\ensuremath{\decompsym{#1}}}
  \newcommand{\algcar}[1]{\ensuremath{\mathalpha{#1}}}
  \newcommand{\mon}[1][a]{\algel{#1}}

    \newcommand{\monx}[1][]{\mon[x_{#1}]}
    \newcommand{\mony}[1][]{\mon[y_{#1}]}
    \newcommand{\monz}[1][]{\mon[z_{#1}]}
  \newcommand{\Mon}[1][M]{\algcar{#1}}
  \newcommand{\prim}[1][p]{\algel{#1}}
    \newcommand{\primp}[1][]{\prim[p_{#1}]}
    \newcommand{\primq}[1][]{\prim[q_{#1}]}

  \newcommand{\emcomp}{\ensuremath{\mathalpha{e}}}
  \newcommand{\compsym}{\ensuremath{\mathbin{\cdot}}}
  \newcommand{\comp}{}
  
   \newcommand{\ord}{\ensuremath{\mathrel{\preccurlyeq}}}
  \newcommand{\invord}{\ensuremath{\mathrel{\succcurlyeq}}}
  \newcommand{\sord}{\ensuremath{\mathrel{\prec}}}
  \newcommand{\invsord}{\ensuremath{\mathrel{\succ}}}
  \newcommand{\mset}[1][m]{\algel{#1}}
    \newcommand{\msetm}{\mset[m]}
    \newcommand{\msetn}{\mset[n]}
  \newcommand{\emptymset}{\ensuremath{\mathalpha{\Box}}}
  \newcommand{\multiplussym}{\ensuremath{\mathalpha{\uplus}}}
  \newcommand{\multiplus}{\ensuremath{\mathbin{\multiplussym}}}
  \newcommand{\multiminsym}{\ensuremath{\mathalpha{-}}}
  \newcommand{\multimin}{\ensuremath{\mathbin{\multiminsym}}}
  \newcommand{\multimaalsym}{\ensuremath{\mathalpha{\cdot}}}
  \newcommand{\multimaal}{\ensuremath{\mathbin{\multimaalsym}}}

  \newcommand{\makemset}[1]{\ensuremath{\lbag{#1}\rbag}}
  
  \newdimen\boxwdplusemdimen
  \def\arrow#1{{
    \boxwdplusemdimen=1em%
    \setbox0=\hbox{$\scriptstyle#1$}%
    \advance\boxwdplusemdimen by \wd0\relax%
    \ifdim\boxwdplusemdimen<16.11119pt%
      \boxwdplusemdimen=16.11119pt%
    \fi%
    \buildrel{#1}\over%
      {\setbox1=\hbox to \boxwdplusemdimen{\rightarrowfill}%
    \ht1=0.3em\relax\box1}%
  }}
  \newcommand{\step}[1]{\ensuremath{\mathbin{\arrow{#1}}}}

   \def\iarrow#1{{
     \boxwdplusemdimen=1em%
     \setbox0=\hbox{$\scriptstyle#1$}%
     \advance\boxwdplusemdimen by \wd0\relax%
     \ifdim\boxwdplusemdimen<16.11119pt%
       \boxwdplusemdimen=16.11119pt%
     \fi%
     \buildrel{#1}\over%
       {\setbox1=\hbox to \boxwdplusemdimen{\leftarrowfill}%
     \ht1=0.3em\relax\box1}%
   }}
  \makeatletter
  \def\twoheadrightarrowfill{$\m@th\smash-\mkern-7mu%
  \cleaders\hbox{$\mkern-2mu\smash-\mkern-2mu$}\hfill
  \mkern-7mu\mathord\twoheadrightarrow$}
\def\darrow#1{{
 \boxwdplusemdimen=1em%
 \setbox0=\hbox{$\scriptstyle#1$}%
 \advance\boxwdplusemdimen by \wd0\relax%
 \ifdim\boxwdplusemdimen<16.11119pt%
   \boxwdplusemdimen=16.11119pt%
 \fi%
 \buildrel{#1}\over%
   {\setbox1=\hbox to \boxwdplusemdimen{\twoheadrightarrowfill}%
     \ht1=0.3em\relax\box1}%
   }}
\makeatother

\newcommand{\ssteps}[1][]{\ensuremath{\mathbin{\darrow{#1}}}}
\def\mdarrow#1{\twoheadrightarrow\mkern-15mu{
 \boxwdplusemdimen=1em%
 \setbox0=\hbox{$\scriptstyle#1$}%
 \advance\boxwdplusemdimen by \wd0\relax%
 \ifdim\boxwdplusemdimen<16.11119pt%
   \boxwdplusemdimen=16.11119pt%
 \fi%
 \buildrel{#1}\over%
   {\setbox1=\hbox to \boxwdplusemdimen{\rightarrowfill}%
     \ht1=0.3em\relax\box1}%
   }}
\newcommand{\silntsym}{\ensuremath{\mathalpha{\rotatebox[origin=c]{-90}{$\twoheadrightarrow$}}}}
\newcommand{\silnt}[1]{\ensuremath{{#1}\mathclose{\silntsym}}}
\newcommand{\opt}[1]{\mbox{\tiny\rm(}#1\mbox{\tiny\rm)}} 
\newcommand{\optstep}[1]{\step{\opt{#1}}}
  \newcommand{\dc}[1][d]{\algel{#1}}
    \newcommand{\dcd}[1][]{\dc[d_{#1}]}

  \newcommand{\dceq}{\ensuremath{\mathrel{\equiv}}}

  \newcommand{\dordsym}{\ensuremath{\mathalpha{\trianglelefteq}}}
  \newcommand{\dord}{\ensuremath{\mathrel{\dordsym}}}
  \newcommand{\sdordsym}{\ensuremath{\mathrel{\vartriangleleft}}}
  \newcommand{\sdord}{\ensuremath{\mathrel{\sdordsym}}}
  \newcommand{\Cdordsym}{\ensuremath{\mathalpha{\trianglelefteq_{\times}}}}
  \newcommand{\Cdord}{\ensuremath{\mathrel{\Cdordsym}}}

\newtheorem{thm}{Theorem}
\newtheorem{cor}[thm]{Corollary}
\newtheorem{lem}[thm]{Lemma}

\newtheorem{prop}[thm]{Proposition}

\theoremstyle{definition}
\newtheorem{defn}[thm]{Definition}
\newtheorem{exmp}[thm]{Example}
\newtheorem{rem}[thm]{Remark}

  \newtheorem{notn}[thm]{Notation}
  \newcommand{\Case}[1]{\item[\normalfont\textsc{Case #1:}]}
  
  \newenvironment{distinction}{\begin{description}}{\end{description}}
  \newcommand{\N}{\ensuremath{\mathbf{N}}}

  \newcommand{\telop}{\ensuremath{\mathbin{+}}}

  \newcommand{\MSet}[1]{\ensuremath{\mathcal{M}(#1)}}

\newcommand{\wn}[1]{\ensuremath{\mathit{n}(#1)}}
\newcommand{\wdepth}[1]{\ensuremath{\mathit{d}(#1)}}

\newcommand{\Figure}[1]{Figure~\ref{fig:#1}}
\newcommand{\Table}[1]{Table~\ref{tab:#1}}
\newcommand{\Lemma}[1]{Lemma~\ref{lem:#1}}
\newcommand{\Proposition}[1]{Proposition~\ref{prop:#1}}
\newcommand{\Definition}[1]{Definition~\ref{def:#1}}
\newcommand{\Theorem}[1]{Theorem~\ref{thm:#1}}
\newcommand{\Corollary}[1]{Corollary~\ref{cor:#1}}
\newcommand{\Example}[1]{Example~\ref{exa:#1}}
\newcommand{\Section}[1]{Section~\ref{sec:#1}}
\newcommand{\Remark}[1]{Remark~\ref{rem:#1}}

\begin{document}
\title{Unique Parallel Decomposition
         in Branching and Weak Bisimulation Semantics}
\author{Bas Luttik}
\date{}

\maketitle

\begin{abstract}
  We consider the property of unique parallel decomposition modulo
  branching and weak bisimilarity. First, we show that normed
  behaviours always have parallel decompositions, but that these are
  not necessarily unique. Then, we establish that finite behaviours
  have unique parallel decompositions. We derive the latter result
  from a general theorem about unique decompositions in partial
  commutative monoids.
\end{abstract}

\section{Introduction}\label{sec:introduction}

A recurring question in process theory is to what extent the
behaviours definable in a certain process calculus admit a unique
decomposition into indecomposable parallel components. Milner and
Moller \cite{MM93} were the first to address the question. They proved
a unique parallel decomposition theorem for a simple process calculus,
which allows the specification of finite behaviour up to strong
bisimilarity and includes parallel composition in the form of pure
interleaving without interaction between the components. They also
presented counterexamples showing that unique parallel decomposition
may fail in process calculi in which it is possible to specify
infinite behaviour, or in which certain coarser notions of behavioural
equivalence are used.

Moller proved several more unique parallel
decomposition results in his dissertation \cite{Mol89},
replacing interleaving parallel composition by CCS parallel
composition, and then also considering weak bisimilarity. These
results were established with subsequent refinements of an ingenious
proof technique attributed to Milner. Christensen, in his
  dissertation \cite{Chr93}, further refined the proof technique to
make it work for the \emph{normed} behaviours recursively definable
modulo strong bisimilarity, and for \emph{all} behaviours recursively
definable modulo distributed bisimilarity.

With each successive refinement of Milner's proof technique, the
technical details became more complicated, but the general idea of the
proof remained the same. In \cite{LO05} we made an attempt
to isolate the deep insights from the technical details, by
identifying a sufficient condition on partial commutative monoids that
facilitates an abstract version of Milner's proof technique.
To concisely present the sufficient condition, we have put forward
the notion of \emph{decomposition order}; it is established in
\cite{LO05}, by means of an abstract version of Milner's technique,
that if a partial commutative monoid can be endowed with a
decomposition order, then it has unique decomposition.

Application of the general result of \cite{LO05} in commutative
monoids of behaviour is often straightforward: a well-founded order
naturally induced on behaviour by (a terminating fragment of) the
transition relation typically satisfies the properties of a
decomposition order. All the aforementioned unique parallel
decomposition results can be directly obtained in this way, except
Moller's result that finite behaviours modulo weak bisimilarity have
unique decomposition. It turns out that a decomposition order cannot
straightforwardly be obtained from the transition relation if certain
transitions are deemed unobservable by the behavioural equivalence
under consideration.

In this article, we address the question of how to establish unique
parallel decomposition in settings with a notion of unobservable
behaviour. Our main contribution will be an adaptation of the general
result in \cite{LO05} to make it suitable for establishing unique
parallel decomposition also in settings with a notion of unobservable
behaviour. To illustrate the result, we shall apply it to establish
unique parallel decomposition for finite behaviour modulo branching or
weak bisimilarity. We shall also show, by means of a counterexample,
that unique parallel decomposition fails for infinite behaviours
modulo branching and weak bisimilarity, even if only a very limited
form of infinite behaviour is considered (normed behaviour
definable in a process calculus with prefix iteration).


A positive answer to the unique parallel decomposition question seems
to be primarily of theoretical interest, as a tool for proving other
theoretical properties about process calculi. For instance, Moller's proofs in
\cite{Mol90b,Mol90a} that PA and CCS cannot be finitely axiomatised
without auxiliary operations and Hirshfeld and Jerrum's proof in
\cite{HJ99} that bisimilarity is decidable for normed PA both rely on
unique parallel decomposition. When parallel composition cannot be
eliminated from terms by means of axioms, then unique parallel
decomposition is generally used to find appropriate normal forms in
completeness proofs for equational axiomatisations
\cite{AFIL05b,AFIL09,AILT08,FL00,HP08}. In \cite{LPSS11}, a
unique parallel decomposition result serves as a stepping stone for
proving complete axiomatisation and decidability results in the
context of a higher-order process calculus.

There is an intimate relationship between unique parallel
decomposition and of cancellation with respect to parallel
composition; the properties are in most circumstances equivalent. In
\cite{CH89}, cancellation with respect parallel composition was first
proved and exploited to prove the completeness of an axiomatisation of
distributed bisimilarity.

Unique parallel decomposition could be of practical interest too,
e.g., to devise methods for finding the maximally parallel
implementation of a behaviour \cite{CGM98}, or for improving
verification methods \cite{GM92}. In \cite{DELL13}, unique parallel
decomposition results are established for the Applied $\pi$-calculus,
as a tool in the comparison of different security notions in the
context of  electronic voting.

This article is organised as follows. In \Section{preliminaries} we
introduce the process calculus that we shall use to illustrate our
theory of unique decomposition. There, we also present counterexamples
to the effect that infinite behaviours in general may not have a
decomposition, and normed behaviours may have more than one
decomposition. In \Section{pcm} we recap the theory of decomposition
put forward in \cite{LO05} and discuss why it is not readily
applicable to establish unique parallel decomposition for finite
behaviours modulo branching and weak
bisimilarity. In \Section{uniqueness} we adapt the theory of
\cite{LO05} to make it suitable for proving unique parallel
decomposition results in process calculi with a notion of
unobservability. In \Section{application} we apply the theorem
from \Section{uniqueness}, showing that bounded behaviours have a
unique parallel decomposition both modulo branching and weak
bisimilarity. We end the article in \Section{conclusions} with a
short conclusion.

  An extended abstract of this article appeared as \cite{Lut12b}.

\section{Processes up to branching and weak bisimilarity}
  \label{sec:preliminaries}

  We define a simple language of process expressions together with an
  operational semantics, and notions of branching and weak
  bisimilarity. We shall then investigate to what extent process
  expressions modulo branching or weak bisimilarity admit parallel
  decompositions. We shall present examples of process expressions
  without a decomposition, and of normed process expressions
  with two distinct decompositions.

  \paragraph{Syntax}

  We fix a set $\Act$ of \emph{actions}, and declare a special action
  $\silent$ that we assume is not in $\Act$.
  We denote by $\Actt$ the set $\Act\cup\{\silent\}$, and
  we let $\acta$ range over $\Act$ and $\act$ over $\Actt$.
  The set $\PTERMS$ of \emph{process expressions} is generated by the
  following grammar:
  \begin{equation*}
    \cptermP ::=\ \nil\ \mid\
                  \pref{\act}\cptermP\ \mid\
                  \cptermP\altc\cptermP\ \mid\
                  \cptermP\merge\cptermP\ \mid\
                  \starpref{\act}\cptermP\
  \qquad\text{($\act\in\Actt$)}.
  \end{equation*}
  The language above is \textsf{BCCS} (the core of Milner's
  \textsf{CCS} \cite{Mil89}) extended with a construction $\_\merge\_$
  to express interleaving parallelism and the prefix iteration
  construction $\starpref{\act}\_$ to specify a restricted form of
  infinite behaviour. We include only a very basic notion of parallel
  composition in our calculus, but note that this is just to simplify
  the presentation. Our unique decomposition theory extends
  straightforwardly to more intricate notions of parallel composition,
  e.g., modelling some form of communication between components.
  To be able to omit some parentheses when writing process
  expressions,   we adopt the conventions that $\pref{\act}$ and
  $\starpref{\act}$   bind stronger, and that $\altc$ binds weaker
  than all the other operations.

  \paragraph{Operational semantics and branching and weak bisimilarity}

  \begin{table} \small
  \center

  \begin{tabular}{c} \\
    $\dfrac{}{\pref{\act}\cptermP\step{\act}\cptermP}$
  \qquad
    $\dfrac{\cptermP\step{\act}\cpterm[P']}
           {\cptermP\altc\cptermQ\step{\act}\cpterm[P']}$
  \qquad
    $\dfrac{\cptermQ\step{\act}\cpterm[Q']}
           {\cptermP\altc\cptermQ\step{\act}\cpterm[Q']}$
\\ \\
    $\dfrac{\cptermP\step{\act}\cpterm[P']}
           {\cptermP\merge\cptermQ\step{\act}\cpterm[P']\merge\cptermQ}$
  \qquad
    $\dfrac{\cptermQ\step{\act}\cpterm[Q']}
           {\cptermP\merge\cptermQ\step{\act}\cptermP\merge\cpterm[Q']}$
\qquad
    $\dfrac{}
           {\starpref{\act}\cptermP\step{\act}\starpref{\act}\cptermP}$
  \qquad
    $\dfrac{\cptermP\step{\beta}\cpterm[P']}
           {\starpref{\act}{\cptermP}\step{\beta}\cpterm[P']}$
  \end{tabular}\bigskip
  \caption{The operational semantics.}
    \label{tab:tss}
  \end{table}
  
  We define on $\CPTERMS$ binary relations $\step{\act}$
  ($\act\in\Actt$) by means of the operational rules in \Table{tss}.
  We denote by $\ssteps$ the reflexive-transitive closure of
  $\step{\silent}$, i.e., 
     $\ptermP{}\ssteps\pterm[P']$
  if there exist $\ptermP[0],\dots,\ptermP[n]$ ($n\geq 0$) such that
    $\ptermP{} =
    \ptermP[0] \step{\silent}\cdots\step{\silent} \ptermP[n]
      =\pterm[P']$.
  Furthermore, we shall write $\ptermP\optstep{\act}\pterm[P']$ if
  $\ptermP\step{\act}\pterm[P']$ or $\act=\silent$ and
  $\ptermP=\pterm[P']$.

  \begin{defn}[Branching bisimilarity \cite{GW96}]\label{def:bbisim}
    A symmetric binary relation $\brelsym$ on $\CPTERMS$ is a
    \emph{branching bisimulation} if for all
    $\cptermP,\cptermQ\in\CPTERMS$ such that $\cptermP\brel\cptermQ$ and for
    all $\act\in\Actt$ it holds that
    \begin{quotation}\noindent
      if
        $\cptermP\step{\act}\cptermP'$ for some $\cptermP'\in\CPTERMS$,
      then
        there exist $\cpterm[Q''],\cpterm[Q']\in\CPTERMS$ such that
          $\cptermQ\ssteps\cpterm[Q'']\optstep{\act}\cpterm[Q']$ and
          $\cptermP\brel\cpterm[Q'']$ and $\cpterm[P']\brel\cpterm[Q']$.
    \end{quotation}
    We write $\cptermP\bbisim\cptermQ$ if there exists a branching
    bisimulation $\brelsym$ such that $\cptermP\brel\cptermQ$.
  \end{defn}

  The relation $\bbisim$ is an equivalence relation on $\CPTERMS$
  (this is not as trivial as one might expect; for a proof see
  \cite{Bas96}). It is also compatible with the construction of
  parallel composition in our syntax, which means that, for all
    $\cptermP[1], \cptermP[2],\cptermQ[1],\cptermQ[2]\in\CPTERMS$:
  \begin{equation} \label{eq:bbisimcompatible}
    \cptermP[1]\bbisim\cptermQ[1]\
  \text{and}\
    \cptermP[2]\bbisim\cptermQ[2]\
  \text{implies}\
     \cptermP[1]\merge\cptermP[2]\bbisim\cptermQ[1]\merge\cptermQ[2]
  \enskip.
  \end{equation}
  (The relation $\bbisim$ is also compatible with $\pref{\alpha}$, but
  not with $\altc$ and $\starpref{\alpha}$. In this article, we shall
  only rely on compatibility with $\merge$.)

\begin{defn}[Weak bisimilarity \cite{Mil90}]\label{def:wbisim}
   A symmetric binary relation $\brelsym$ on $\CPTERMS$ is a
   \emph{weak bisimulation} if for  all $\cptermP,\cptermQ\in\CPTERMS$
   such that $\cptermP\brel\cptermQ$ and for all $\act\in\Actt$ it holds that
    \begin{quotation}\noindent
      if
        $\cptermP\step{\act}\cptermP'$ for some $\cptermP'\in\CPTERMS$,
      then
        there exist
        $\cpterm[Q'],\cpterm[Q''],\cpterm[Q''']\in\CPTERMS$
      such that
          $\cptermQ\ssteps\cpterm[Q'']\optstep{\act}\cpterm[Q''']\ssteps\cpterm[Q']$
          and $\cpterm[P']\brel\cpterm[Q']$.
    \end{quotation}
    We write $\cptermP\wbisim\cptermQ$ if there exists a weak
    bisimulation $\brelsym$ such that $\cptermP\brel\cptermQ$.
  \end{defn}

  The relation $\wbisim$ is an equivalence relation on $\CPTERMS$. It
  is also compatible with parallel composition, i.e., for all
    $\cptermP[1], \cptermP[2],\cptermQ[1],\cptermQ[2]\in\CPTERMS$:
  \begin{equation} \label{eq:wbisimcompatible}
    \cptermP[1]\wbisim\cptermQ[1]\
  \text{and}\
    \cptermP[2]\wbisim\cptermQ[2]\
  \text{implies}\
     \cptermP[1]\merge\cptermP[2]\wbisim\cptermQ[1]\merge\cptermQ[2]
  \enskip.
  \end{equation}
  (Just like $\bbisim$, the relation $\wbisim$ is not compatible
  with $\altc$ and $\starpref{\alpha}$.)
  Note that $\bbisim\subseteq\wbisim$; we shall often implicitly use
  this property below.

A process expression is \emph{indecomposable} if it is not
behaviourally equivalent to $\nil$ or a non-trivial parallel
composition (a parallel composition is trivial if one of its
components is behaviourally equivalent to $\nil$). We say that a
process theory has \emph{unique parallel decomposition} if every
process expression is behaviourally equivalent to a unique
(generalised) parallel composition of indecomposable process
expressions. Uniqueness means that the indecomposables of any two
decompositions of a process expression are pairwise behaviourally
equivalent up to a permutation.

We should make the definitions of indecomposable and unique parallel
decomposition more formal and concrete for the two behavioural
equivalences considered in this article (viz.\ branching and weak bisimilarity). For reasons of
generality and succinctness, however, it is convenient to postpone our
formalisation until the next section, where we will discuss
decomposition in the more abstract setting of commutative monoids. For
now, we rely on the intuition of the reader and discuss informally and
by means of examples to what extent the process theory
introduced above might have the property of unique parallel
decomposition. In our explanations we use branching bisimilarity as
behavioural equivalence, but everything we say in the remainder of
this section remains valid if branching bisimilarity is replaced by
weak bisimilarity.

The first observation, already put forward by Milner and Moller in
\cite{MM93}, is that there are process expressions that do not have a
decomposition at all. (In \cite{MM93}, the following example is
actually used to show that there exist infinite processes which do not
have a decomposition modulo \emph{strong} bisimilarity.)
\begin{exmp} \label{exa:nodecompositions}
  Consider the process expression
    $\starpref{\acta}\nil$ (with $\acta\not=\silent$),
  and suppose that $\starpref{\acta}\nil$ has a decomposition, say
  $\starpref{\acta}\nil\bbisim\cptermP\merge\cptermQ$ for some process
  expressions $\cptermP$ and $\cptermQ$.

  We first argue that either
    $\ptermP\bbisim\starpref{\acta}\nil$
  or
    $\ptermQ\bbisim\starpref{\acta}\nil$.
  Note that it follows from
    $\starpref{\acta}\nil\bbisim\cptermP\merge\cptermQ$
  that all process expressions reachable from
  $\cptermP\merge\cptermQ$ are branching bisimilar to
    $\starpref{\acta}\nil$,
  and hence for all process expressions $\cptermR$ reachable from
  $\cptermP$ we have that $\cptermR\step{\act}\cptermR'$ 
  implies $\act=\acta$. So if $\cptermR$ is reachable from $\cptermP$,
  then either $\cptermR\bbisim\nil$ or there exists $\cptermR'$ such
  that $\cptermR\step{\acta}\cptermR'$.
  If, on the one hand, there exists a process expression $\cptermR$
  reachable from $\cptermP$ such that $\cptermR\bbisim\nil$, then, since
  $\cptermR\merge\cptermQ$ is reachable from $\cptermP\merge\cptermQ$,
  it follows that
    $\starpref{\acta}\nil\bbisim\cptermR\merge\cptermQ\bbisim\cptermQ$.
  If, on the other hand, there does not exist a process expression
  $\cptermR$ reachable from $\cptermP$ such that
  $\cptermR\bbisim\nil$, then, for all process expressions $\cptermR$
  reachable from $\cptermP$, on the one hand there exists $\cptermR'$
  such that $\cptermR\step{\acta}\cptermR'$ and on the other hand for
  all $\cptermR'$ and $\act$ such that $\cptermR\step{\act}\cptermR'$
  we have that $\act=\acta$. Therefore, the relation
    $\brelsym =
         \{ (\starpref{\acta}\nil,\cptermR) \mid
                 \text{$\cptermR$ is reachable from $\cptermP$}\}$
  is a (branching) bisimulation relation, and hence
    $\starpref{\acta}\nil\bbisim\cptermP$.

  Now, since
    $\starpref{\acta}\nil\bbisim\cptermP\merge\cptermQ$
  implies that either
    $\ptermP\bbisim\starpref{\acta}\nil$
  or
    $\ptermQ\bbisim\starpref{\acta}\nil$,
  a decomposition of $\starpref{\acta}\nil$ would necessarily include
  an indecomposable branching bisimilar to $\starpref{\acta}\nil$.
  But, since
     $\starpref{\acta}\nil 
         \bbisim
      \starpref{\acta}\nil\merge\starpref{\acta}\nil$,
  there does not exist an indecomposable branching bisimilar to
  $\starpref{\acta}\nil$. We conclude that $\starpref{\acta}\nil$
  fails to have a decomposition.
\end{exmp}

Note that the process expression $\starpref{\acta}\nil$ does not admit
terminating behaviour; it does not have a transition sequence to a
process expression from which no further transitions are possible.  We
want to identify a conveniently large subset of process expressions
that do have decompositions, and to exclude the counterexample against
existence of decompositions, we confine our attention to process
expressions with terminating behaviour.

For $\acta\in\Act$ and process expressions $\cptermP$ and $\cptermQ$
we write $\cptermP\ssteps[\acta]{}\cptermQ$ whenever there exist
process expressions $\cptermP'$ and $\cptermQ'$ such that
  $\cptermP\ssteps{}\cptermP'\step{\acta}\cptermQ'\ssteps{}\cptermQ$.
We say that $\ptermP$ is \emph{silent} and write $\silnt{\ptermP}$ if
there do not exist $\acta\in\Act$ and $\ptermQ$ such that $\ptermP\ssteps[{\acta}]\ptermQ$.
  
\begin{defn}
  A process expression $\cptermP$ is \emph{normed}
  if there exist
    a natural number $k\in\N$,
    process expressions $\ptermP[0],\dots,\ptermP[k]\in\PTERMS$
  and
    actions $\acta[1],\dots,\acta[k]\in\Act$
  such that
     $\cptermP=\cptermP[0]
                       \ssteps[{\acta[1]}]{}\cdots\ssteps[{\acta[k]}]{}
                    \cptermP[k]$
  and $\silnt{\cptermP[k]}$.
  The \emph{norm} $\wn{\cptermP}$ of a normed process
  expression $\cptermP$ is defined by
\begin{equation*}
  \wn{\cptermP}=\min
    \{k: \is{\cptermP[0],\dots,\cptermP[k]\in\PTERMS}.\
                  \is{\acta[1],\dots,\acta[k]\in\Act}.\
                     \cptermP=\cptermP[0]
                       \ssteps[{\acta[1]}]{}\cdots\ssteps[{\acta[k]}]{}
                    \silnt{\cptermP[k]}\}
\enskip.
\end{equation*}
\end{defn}

It is immediate from their definitions that both branching and weak
bisimilarity preserve norm: if two process expressions 
are branching or weakly bisimilar, then they have equal norms.
It is also easy to establish that a parallel composition is
normed if, and only if, both parallel components are normed. In fact,
norm is additive with respect to parallel composition: the norm of a
parallel composition is the sum of the norms of its parallel
components. Note that a process expression with norm $0$ is
behaviourally equivalent to $\nil$.

With a straightforward induction on norm it can be established
that normed process expressions have a decomposition. But
sometimes even more than one, as is illustrated in the following
example. 
\begin{exmp} \label{exa:wninsufficient}
  Consider the process expressions
    $\cptermP=\starpref{\acta}\pref{\silent}\pref{\actb}\nil$
  and
    $\cptermQ=\pref{\actb}\nil$.
  It is clear that $\cptermP$ and $\cptermQ$ are \emph{not}
  branching bisimilar.
  Both $\cptermP$ and $\cptermQ$ have norm $1$, and from this
  it immediately follows that they are both indecomposable.
  Note that, according to the operational semantics,
  $\cptermP\merge\cptermP$ gives rise to the following three
  transitions:
  \begin{enumerate}
  \item $\cptermP\merge\cptermP\step{\acta}\cptermP\merge\cptermP$;
  \item $\cptermP\merge\cptermP\step{\silent}\cptermP\merge\cptermQ$;
    and
  \item $\cptermP\merge\cptermP\step{\silent}\cptermQ\merge\cptermP$.
  \end{enumerate}
  Further note that
  $\cptermP\merge\cptermQ\step{\acta}\cptermP\merge\cptermQ$ and
  $\cptermQ\merge\cptermP\step{\acta}\cptermQ\merge\cptermP$.
  (The complete transition graph associated with
  $\cptermP\merge\cptermP$ by the operational semantics is shown in
  \Figure{counterexample}.)
  Using these facts it is straightforward to verify that the reflexive-symmetric
  closure of the binary relation
  \begin{multline*}
     \brelsym{} =
        \{ (\cptermP\merge\cptermP,\cptermP\merge \cptermQ),
            (\cptermP\merge\cptermP,\cptermQ\merge\cptermP)
        \} \\
       \mbox{}\cup
       \{  (\cptermP\merge\cptermQ,\cptermQ\merge\cptermP),
            (\cptermP\merge\nil,\nil\merge\cptermP),
            (\cptermQ\merge\nil,\nil\merge\cptermQ)
       \}
  \end{multline*}
  is a branching bisimulation, and hence
    $\cptermP\merge\cptermP
  \bbisim
    \cptermP\merge\cptermQ$.
   It follows that $\cptermP\merge\cptermP$ and
   $\cptermP\merge\cptermQ$ are distinct decompositions of the same
   process up to branching bisimilarity.
  \end{exmp}

\begin{figure}
    \begin{center}
     \begin{transsys}(35,35)
      \graphset{iangle=90,fangle=-90}
      {\small
      \node[Nmarks=i](s0)(0,30){$\cptermP\merge\cptermP$}
      \node(s1)(20,30){$\cptermP\merge\cptermQ$}
      \node(s2)(40,30){$\cptermP\merge\nil$}
      \node(s3)(0,15){$\cptermQ\merge\cptermP$}
      \node(s4)(20,15){$\cptermQ\merge\cptermQ$}
      \node(s5)(40,15){$\cptermQ\merge\nil$}
      \node(s6)(0,0){$\nil\merge\cptermP$}
      \node(s7)(20,0){$\nil\merge\cptermQ$}
      \node(s8)(40,0){$\nil\merge\nil$}
       \edge*[loopangle=180,loopdiam=6](s0){$\acta$}
       \edge*[loopangle=180,loopdiam=6](s3){$\acta$}
       \edge*[loopangle=180,loopdiam=6](s6){$\acta$}
       \edge*[loopangle=90,loopdiam=6](s1){$\acta$}
       \edge*[loopangle=90,loopdiam=6](s2){$\acta$}
       \edge(s0,s1){$\silent$}
       \edge(s1,s2){$\actb$}
       \edge(s0,s3){$\silent$}
       \edge(s1,s4){$\silent$}
       \edge(s2,s5){$\silent$}
       \edge(s3,s4){$\silent$}
       \edge(s4,s5){$\actb$}
       \edge(s4,s7){$\actb$}
       \edge(s5,s8){$\actb$}
       \edge(s3,s6){$\actb$}
       \edge(s6,s7){$\silent$}
       \edge(s7,s8){$\actb$}
      }
     \end{transsys}
    \end{center}
   \caption{Transition graph associated with $\cptermP\merge\cptermP$.}\label{fig:counterexample}
\end{figure}

Incidentally, the processes in the above counterexample also refute
claims in \cite{FL09} to the effect that processes definable with a
normed $\mathsf{BPP}$ specification have a unique
decomposition modulo branching bisimilarity and weak bisimilarity.

  Apparently, more severe restrictions are needed.

\begin{defn}
  Let $k\in\N$; a process expression $\cptermP$ is \emph{bounded} by
  $k$ if for all $\ell\in\N$ the existence of
    $\cptermP[1],\dots,\cptermP[\ell]\in\PTERMS$
  and
    $\acta[1],\dots,\acta[\ell]\in\Act$
  such that
    $\cptermP\ssteps[{\acta[1]}]\cdots\ssteps[{\acta[\ell]}]\cptermP[\ell]$
  implies that $\ell\leq k$.
  We say that $\cptermP$ is \emph{bounded} if $\cptermP$ is
  bounded by $k$ for some $k\in\N$.
\end{defn}

The \emph{depth} $\wdepth{\cptermP}$ of a bounded process
expression $\cptermP$ is the length of its longest transition sequence
not counting $\silent$-transitions, i.e., 
  \begin{equation*}
    \wdepth{\cptermP}=
       \max\{k: \is{\cptermP[0],\dots,\cptermP[k]}.\
                      \is{\acta[1],\dots,\acta[k]}.\
                          \cptermP=\cptermP[0]
                            \ssteps[{\acta[1]}]\cdots\ssteps[{\acta[k]}]
                          \cptermP[k]
       \}
  \enskip.
  \end{equation*}

Both branching and weak bisimilarity preserve depth: if two process
expressions are branching or weakly bisimilar, then they have equal
depth. Furthermore, depth is additive with respect to parallel
composition: the depth of a parallel composition is the sum of the
depths of its parallel components. A process expression with depth $0$
is behaviourally equivalent to $\nil{}$.

In the remainder of this article we shall establish that bounded
process expressions have a unique parallel decomposition both modulo
branching and weak bisimilarity. We shall derive these results from a
more general result about unique decomposition in commutative monoids.

\section{Partial commutative monoids and
  decomposition} \label{sec:pcm}

In this section we recall the abstract algebraic notion of partial
commutative monoid, and formulate the property of unique
decomposition. We shall see that the process theories discussed in the
previous section give rise to commutative monoids of processes with
parallel composition as binary operation. The notion of unique
decomposition associated with these commutative monoids coincides with
the notion of unique parallel decomposition as discussed.

Then, we shall recall the notion of decomposition order on partial
commutative monoids proposed in \cite{LO05}.  We shall investigate
whether the notion of decomposition order can be employed to prove
unique parallel decomposition of bounded process expressions
modulo branching and weak bisimilarity.

\begin{defn}[{\cite{LO05}, Definition 1}] \label{def:pcm}
\renewcommand{\comp}{\ensuremath{\mathrel{\compsym}}}
  A \emph{(partial) commutative monoid} is a set $\Mon$ with a
  distinguished element $\emcomp$ and a (partial) binary operation on
  $\Mon$ (for clarity in this definition denoted by $\comp$) such that
  for all $\monx, \mony, \monz\in\Mon$:
  \begin{alignat*}{2}
  &  \monx\comp(\mony\comp\monz)\pequate(\monx\comp\mony)\comp\monz
      &&\qquad\text{(associativity)};\\
  &  \monx\comp\mony\pequate\mony\comp\monx
      &&\qquad\text{(commutativity)};\\
  &  \monx\comp\emcomp\pequate\emcomp\comp\monx\pequate\monx
      &&\qquad\text{(identity)}.
  \end{alignat*}
  (The symbol $\pequate$ expresses that either both sides of the
  equation are undefined or both sides are defined and designate the
  same element in $\Mon$; see Remark 2 in \cite{LO05} for further 
  explanations.)
\end{defn}
The symbol $\compsym$ will be omitted if this is unlikely to cause
confusion. Also, we shall sometimes use other symbols ($\merge$, $+$,
$\dots$) to denote the binary operation of a partial commutative
monoid.

  In \cite{LO05}, the key notions of the general theory of
  decomposition for commutative monoids are illustrated using three
  examples: the commutative monoid of natural numbers with addition,
  the commutative monoid of positive natural numbers with
  multiplication, and the commutative monoid of multisets over some
  set. Here we recap the latter example, because we need some of the
  definitions pertaining to multisets in the remainder of this article.

  \begin{exmp} \label{exa:pcm}
    Let $X$ be any set.
    A \emph{(finite) multiset} over $X$ is a mapping
      $\msetm: X \rightarrow \N$
    such that $\msetm(x)>0$ for at most finitely many $x\in X$;
    the number $\msetm(x)$ is called the \emph{multiplicity} of $x$ in
    $\msetm$.
    The set of all multisets over $X$ is denoted by $\MSet{X}$. If
    $\msetm$ and $\msetn$ are multisets, then their sum
    $\msetm\multiplus\msetn$ is obtained by coordinatewise addition of
    multiplicities, i.e.,
      $(\msetm\multiplus\msetn)(x)=\msetm(x)\telop\msetn(x)$
        for all $x\in X$.
    The \emph{empty multiset} $\emptymset$ is the multiset that
    satisfies $\emptymset(x)=0$ for all $x\in X$.
    With these definitions, $\MSet{X}$ is a commutative monoid. If
    $x_1,\dots,x_k$ is a sequence of elements of $X$, then
    $\makemset{x_1,\dots,x_k}$ denotes the multiset $\msetm$ such that
    $\msetm(x)$ is the number of occurrences of $x$ in
    $x_1,\dots,x_k$.
  \end{exmp}

Process expressions modulo branching or weak bisimilarity also give
rise to commutative monoids. Recall that $\bbisim$ and $\wbisim$ are
equivalence relations on the set of process expressions. We denote the
equivalence class of a process expression $\cpterm$ modulo $\bbisim$
or $\wbisim$, respectively, by $\beqclass{\cpterm}$ and
$\weqclass{\cpterm}$, i.e.,
  \begin{gather*}
    \beqclass{\cpterm}=
      \{\cptermQ\in\CPTERMS:\cptermP\bbisim\cptermQ\}
\enskip;\
      \text{and}\\
    \weqclass{\cpterm}=
      \{\cptermQ\in\CPTERMS:\cptermP\wbisim\cptermQ\}\enskip.
  \end{gather*}
  Then, we define
  \begin{gather*}
     \BPROC
        =\facalg{\CPTERMS}{\bbisim}
        =\{\beqclass{\cpterm}:\cpterm\in\CPTERMS\}
\enskip;\ \text{and}\\
     \WPROC
        =\facalg{\CPTERMS}{\wbisim}
        =\{\weqclass{\cpterm}:\cpterm\in\CPTERMS\}
  \enskip.
  \end{gather*}

  In this article, the similarities between the commutative monoids
  $\BPROC$ and $\WPROC$ will be more important than the differences.
  It will often be necessary to define notions for both commutative
  monoids, in a very similar way. For succinctness of 
  presentation, we allow ourselve a slight \emph{abus de language} and
  most of the time deliberately omit the subscripts $b$ and $w$ from
  our notation for equivalence classes. Thus, we will be able to
  efficiently define notions and prove facts simultaneously for
  $\BPROC$ and $\WPROC$.

  For example, since both $\bbisim$ and $\wbisim$ are compatible with
  $\merge$ (see Equations~\eqref{eq:bbisimcompatible} and
  \eqref{eq:wbisimcompatible}), we can define a binary operation
  $\merge$ simultaneously on $\BPROC$ and $\WPROC$ simply by
\begin{equation*}
    \eqclass{\cptermP}\merge\eqclass{\cptermQ}
  = \eqclass{\cptermP\merge\cptermQ}
  \enskip.
  \end{equation*}
  Also, we agree to write just $\nil$ for $\eqclass{\nil}$.
  It is then straightforward to establish that the binary operation
  $\merge$ is commutative and associative (both on $\BPROC$ and
  $\WPROC$),  and that $\nil$ is the identity element for $\merge$.

  \begin{prop}
    The sets $\BPROC$ and $\WPROC$ are commutative monoids under
    $\merge$, with $\nil$ as identity element.
  \end{prop}

  Note that, since both branching and weak bisimilarity preserve
  depth (norm), whenever an equivalence class $\eqclass{\cptermP}$
  contains a bounded (normed) process expression, it consists entirely
  of bounded (normed) process expressions.
  We define subsets
    $\BPROCfin\subseteq\BPROCtn\subseteq\BPROC$
  and
    $\WPROCfin\subseteq\WPROCtn\subseteq\WPROC$
  by
  \begin{gather*}
     \BPROCfin
        =\{\beqclass{\cpterm}:\cpterm\in\CPTERMS\ \&\ \text{$\cptermP$
          is bounded}\}
  \enskip;\\
     \BPROCtn
        =\{\beqclass{\cpterm}:\cpterm\in\CPTERMS\ \&\ \text{$\cptermP$
          is normed}\}
  \enskip;\\
     \WPROCfin
        =\{\weqclass{\cpterm}:\cpterm\in\CPTERMS\ \&\ \text{$\cptermP$
          is bounded}\}
  \enskip;\ \text{and}\\
     \WPROCtn
        =\{\weqclass{\cpterm}:\cpterm\in\CPTERMS\ \&\ \text{$\cptermP$
          is normed}\}
  \enskip.
  \end{gather*}
  Since a parallel composition is bounded (normed) if, and only if, its
  parallel components are bounded (normed), $\BPROCfin$ and $\BPROCtn$
  are commutative submonoids of $\BPROC$, and $\WPROCfin$ and
  $\WPROCtn$ are commutative submonoids of $\WPROC$.

 \begin{notn}
  Let $\monx_1,\dots,\monx_k$ be a (possibly empty) sequence of
  elements of a monoid $\Mon$; we define its \emph{generalised
    composition} $\monx_1\comp\cdots\comp\monx_k$ as usual.
  Furthermore, we write $\monx^n$ for the $k$-fold composition of
  $\monx$. For explicit definitions of these notations
    see Notation 4 of \cite{LO05}.
  \end{notn}
  It is straightforward by induction to establish the following
  \emph{generalised associative law}:
  \begin{equation*}
    (\monx_1\comp\cdots\comp\monx_k)
       \comp(\mony_1\comp\cdots\comp\mony_{\ell})
  \pequate
    \monx_1\comp\cdots\comp\monx_k\comp\mony_1\comp\cdots\comp\mony_{\ell}
  \enskip.
  \end{equation*}
  Also by induction, a \emph{generalised commutative law} can be
  established, so
  \begin{equation*}
    \text{if $i_1,\dots,i_{\ell}$ is any permutation of $1,\dots,\ell$, then}\
    \monx_1\comp\cdots\comp\monx_{\ell}
  \pequate
    \monx_{i_1}\comp\cdots\comp\monx_{i_{\ell}}
  \enskip.
  \end{equation*}

  An indecomposable element of a commutative monoid is an
  element that cannot be written as a product of two elements that are
  both not the identity element of the monoid.

  \begin{defn}[{\cite{LO05}}, Definition 5] \label{def:indecomposable}
    An element $\primp$ of a commutative monoid $\Mon$ is
    called \emph{indecomposable} if
      $\primp\not=\emcomp$ and
      $\primp=\monx\comp\mony$ implies $\monx=\emcomp$ or
      $\mony=\emcomp$.
  \end{defn}

  \begin{exmp} \label{exa:indec}
\begin{enumerate}
  \item \label{exa:indecmset}
    The indecomposable elements of $\MSet{X}$ are the \emph{singleton}
    multisets, i.e., the multisets $\msetm$ for which it holds that
    $\sum_{x\in X}\msetm(x)=1$.
  \item \label{exa:indecproc}
    The indecomposable elements of $\BPROCfin$, $\BPROCtn$, $\BPROC$,
    $\WPROCfin$, $\WPROCtn$, and $\WPROC$ are the equivalence classes
    of process expressions that are not behaviourally equivalent to
    $\nil$ or a non-trivial parallel composition.
\end{enumerate}
  \end{exmp}

  We define a decomposition in a partial commutative monoid to be a
  finite multiset of indecomposable elements. Note that this gives the
  right notion of equivalence on decompositions, for two finite
  multisets $\makemset{\monx_1,\dots,\monx_k}$ and
  $\makemset{\mony_1,\dots,\mony_{\ell}}$ are equal \IFF{}
  the sequence $\mony_1,\dots,\mony_{\ell}$ can be obtained from the
  sequence $\monx_1,\dots,\monx_k$ by a permutation of its elements.

  \begin{defn}[{\cite{LO05}}, Definition 7]
    Let $\Mon$ be a partial commutative monoid.
    A \emph{decomposition} in $\Mon$ is a finite multiset
      $\makemset{\primp_1,\dots,\primp_k}$
    of indecomposable elements of $\Mon$ such that
    $\primp_1\comp\cdots\comp\primp_k$ is defined.
    The element $\primp_1\comp\cdots\comp\primp_k$ in $\Mon$ will be
    called the \emph{composition} associated with the decomposition
      $\makemset{\primp_1,\dots,\primp_k}$, and, conversely, we say that
      $\makemset{\primp_1,\dots,\primp_k}$
    is a decomposition of the element
      $\primp_1\comp\cdots\comp\primp_k$
    of $\Mon$.
    Decompositions
      $\dcd=\makemset{\primp_1,\dots,\primp_k}$
    and
      $\dc[d']=\makemset{\primp_1',\dots,\primp_{\ell}'}$
    are \emph{equivalent} in $\Mon$ (notation:
    $\dcd\dceq\dc[d']$) if they have the same compositions, i.e., if
    \begin{equation*}
      \primp_1\comp\cdots\comp\primp_k =
      \primp_1'\comp\cdots\comp\primp_{\ell}'
    \enskip.
    \end{equation*}
    A decomposition $\dcd$ in $\Mon$ is \emph{unique} if
      $\dcd\dceq\dc[d']$ implies $\dcd=\dc[d']$
    for all decompositions $\dc[d']$ in $\Mon$.
    We say that an element $\monx$ of $\Mon$ has a unique
    decomposition if it has a decomposition and this decomposition is
    unique; we shall then denote the unique decomposition of $\monx$ by $\decomp{\monx}$.
    If every element of $\Mon$ has a unique decomposition, then we say
    that $\Mon$ has \emph{unique decomposition}.
  \end{defn}

  \begin{exmp}
    Every finite multiset $\msetm$ over $X$ has a unique decomposition
    in $\MSet{X}$, which contains for every $x\in X$ precisely
    $\msetm(x)$ copies of the singleton multiset $\makemset{x}$.
  \end{exmp}

The general notion of unique decomposition for commutative monoids,
when instantiated to one of the commutative monoids of processes
considered in this article, indeed coincides with the notion of unique
parallel decomposition as discussed in the preceding section. We have
already seen that the commutative monoids $\BPROCtn$, $\BPROC$,
$\WPROCtn$ and $\WPROC$ do not have unique decomposition. Our goal
in the remainder of this article is to establish that the commutative
monoids $\BPROCfin$ and $\WPROCfin$ do have unique decomposition.

Preferably, we would like to have a general sufficient condition on
partial commutative monoids for unique decomposition that is easily
seen to hold for $\BPROCfin$ and $\WPROCfin$, and hopefully also for
other commutative monoids of processes. We shall now first recall the
sufficient criterion put forward in \cite{LO05}, which was
specifically designed for commutative monoids of processes. Then, we
shall explain that it cannot directly be applied to conclude that
$\BPROCfin$ and $\WPROCfin$ have unique decomposition. In the next
section, we shall subsequently modify the condition, so that it
becomes applicable to the commutative monoids at hand.

\begin{defn}[\cite{LO05}, Definition 20] \label{def:decomporder}
  Let $\Mon$ be a partial commutative monoid; a partial order $\ord$
  on $\Mon$ is a \emph{decomposition order} if
  \begin{enumerate}
  \renewcommand{\theenumi}{\roman{enumi}}
  \renewcommand{\labelenumi}{(\theenumi)}
    \item \label{def:wf}
      it is \emph{well-founded}, i.e., every non-empty subset of
      $\Mon$ has a $\ord$-minimal element;
    \item \label{def:leastelt}
      the identity element $\emcomp$ of $\Mon$ is the \emph{least element}
      of $\Mon$ with respect to $\ord$, i.e., $\emcomp\ord\monx$ for
      all $\monx$ in $\Mon$;
    \item \label{def:scompatible}
      it is \emph{strictly compatible}, i.e., for
      all $\monx,\mony,\monz\in\Mon$
      \begin{equation*}
        \text{if $\monx\sord\mony$ and $\mony\comp\monz$ is defined,
                 then $\monx\comp\monz\sord\mony\comp\monz$};
      \end{equation*}
    \item \label{def:precomp}
      it is \emph{precompositional}, i.e., for all
      $\monx,\mony,\monz\in\Mon$
      \begin{equation*}
         \text{$\monx\ord\mony\comp\monz$ implies
          $\monx=\mony'\comp\monz'$ for some $\mony'\ord\mony$ and
          $\monz'\ord\monz$};\ \text{and}
      \end{equation*}
    \item \label{def:archimedean}
      it is \emph{Archimedean}, i.e., for all $\monx,\mony\in\Mon$
      \begin{equation*}
        \text{$\monx^n \ord \mony$ for all $n\in\N$ implies that
          $\monx=\emcomp$}.
      \end{equation*}
  \end{enumerate}
\end{defn}

\begin{rem} \label{rem:archimedean}
  In \cite{LO05} a slightly weaker form of the Archimedean property
  (condition (\ref{def:archimedean}) of \Definition{decomporder}) was
  used. In the context of strict compatibility the weaker form was
  enough to arrive at a sufficient condition for unique decomposition
  in partial commutative monoids. We include the stronger version
  here, because we will need to relax the requirement of strict
  compatibility to just compatibility to facilitate application of our
  result in the present setting of weak behavioural equivalences.
\end{rem}

In \cite{LO05} it was proved that the existence of a decomposition
order on a partial commutative monoid is a necessary and sufficient
condition for unique decomposition. The advantage of establishing unique
decomposition via a decomposition order is that it circumvents first
establishing cancellation, which in some cases is hard without knowing
that the partial commutative monoid has unique decomposition. We refer
to \cite{LO05} for a more in-depth discussion.

In commutative monoids of processes, an obvious candidate
decomposition order is the order induced on the commutative monoid by
the transition relation. We define a binary relation $\step{}$ on
$\BPROC$ and $\WPROC$ by
\begin{equation*}
  \eqclass{\cptermP}\step{}\eqclass{\cptermP'}\
     \text{if there exist $\cptermQ\in\eqclass{\cptermP}$,
       $\cptermQ'\in\eqclass{\cptermP'}$ and $\act\in\Actt$
         such that $\cptermQ\step{\act}\cptermQ'$}
\enskip.
\end{equation*}

We shall denote the inverse of the reflexive-transitive closure of
$\step{}$ (both on $\BPROC$ and $\WPROC$) by
$\ord$, i.e., ${\ord}=(\step{}^{*})^{-1}$.

\begin{lem} \label{lem:stepwellbehaved}
  If $\cptermP$ and $\cptermQ$ are process expressions such that
  $\eqclass{\cptermQ}\ord\eqclass{\cptermP}$, then for all
  $\cptermP'\in\eqclass{\cptermP}$ there exist
  $\cptermQ'\in\eqclass{\cptermQ}$,
  $k\geq 0$,
  $\cptermQ[0],\dots,\cptermQ[k]\in\CPTERMS$
and
  $\act_0,\dots,\act_k\in\Actt$
  such that
  \begin{equation*}
    \cptermP'=\cptermQ[0]
      \step{\act_0}\cdots\step{\act_k}
    \cptermQ[k]=\cptermQ'
  \enskip.
  \end{equation*}
\end{lem}

The following lemma implies that every set of process expressions has
minimal elements with respect to the reflexive-transitive closure of
the transition relation. Caution: the lemma holds true only thanks to
the very limited facility for defining infinite behaviour in our
calculus (see also \Remark{infbehaviour} below).

\begin{lem} \label{lem:wellfounded}
  If $\cptermP[0],\dots,\cptermP[i],\dots$ ($i\in\N$) is an infinite
  sequence of process expressions, and $\act_0,\dots,\act_i,\dots$
  ($i\in\N$) is an infinite sequence of elements in $\Actt$ such that
    $\cptermP[i]\step{\act_i}\cptermP[i+1]$ for all $i\in\N$,
  then there exists $j\in\N$ such that
    $\cptermP[k]=\cptermP[\ell]$
  for all $k,\ell\geq j$.
\end{lem}
\begin{proof}
  Define the size $|\cptermP|$ of a process expression $\cptermP$ as the
  number of symbols in $\cptermP$.
  It can then be shown with a straightforward induction on a
  derivation of the transition $\cptermP\step{\act}\cptermP'$
  according to the operational semantics in \Table{tss} that either
  $|\cptermP|>|\cptermP'|$ or $\cptermP=\cptermP'$. From this, the
  lemma clearly follows.
\end{proof}

Using Lemmas~\ref{lem:stepwellbehaved} and \ref{lem:wellfounded} it is
straightforward to establish the following proposition.
\begin{prop} \label{prop:ordgeneral}
  The relation $\ord$ is a well-founded precompositional
  partial order on each of the commutative monoids $\BPROC$,
  $\BPROCtn$, $\BPROCfin$, $\WPROC$, $\WPROCtn$, and $\WPROCfin$.
\end{prop}

\begin{rem}\label{rem:infbehaviour}
  That \Lemma{wellfounded} holds true of our particular process
  calculus, and that, as a consequence (see the following
  proposition), $\ord$ is well-founded on the unrestricted commutative
  monoids $\BPROC$ and $\WPROC$ is thanks to the very limited facility of defining infinite behaviour,
  by means of simple loops. In calculi with more expressive facilities
  to specify infinite behaviour (e.g., recursion, general iteration or
  replication) $\ord$ as defined above is not well-founded (it is not
  even anti-symmetric). Note that the contribution of this article
  does not depend on the well-foundedness of $\BPROC$ and $\WPROC$,
  which is stated for completeness sake.

  It is often possible to define a well-founded partial order on
  processes based on the transition relation in a setting with a more
  general form of infinite behaviour, at least for normed
  processes. See, e.g., \cite{LO05} for an example of an
  anti-symmetric and well-founded order on normed processes definable
  in ACP with recursion, which is based on a restriction of the
  transition relation.
\end{rem}

The ordering $\ord$ defined on $\BPROCtn$, $\BPROC$, $\WPROCtn$ and
$\WPROC$ is not a decomposition order: on $\BPROC$ and $\WPROC$ it
does not satisfy conditions (\ref{def:leastelt}),
(\ref{def:scompatible}) and (\ref{def:archimedean}) of
\Definition{decomporder}, and on $\BPROCtn$ and $\WPROCtn$ it does not
satisfy condition (\ref{def:scompatible}) of
\Definition{decomporder}.
\begin{exmp} \label{exa:nodecompord}
\begin{enumerate}
\item \label{exa:nodecompord:item:nilnotleast}
   Since $\starpref{\acta}\nil\step{\acta}\starpref{\acta}\nil$ is the
   only transition from $\starpref{\acta}\nil$, it follows that
   $\eqclass{\starpref{\acta}\nil}$ is a minimal element of $\BPROC$
   and $\WPROC$ with respect to $\ord$. It is also clear that $\eqclass{\starpref{\acta}\nil}\neq\nil$, so we have that $\nil$ is not the least element of $\ord$ in $\BPROC$ and $\WPROC$.
\item \label{exa:nodecompord:item:notarchimedean}
  In \Example{nodecompositions} we have argued that
  $\starpref{\acta}\nil=\starpref{\acta}\nil\merge\starpref{\acta}\nil$,
  from which it easily follows that
    $\eqclass{\starpref{\acta}\nil}^n=\eqclass{\starpref{\acta}\nil}$
  for all $n\in\N$. Hence, $\ord$ on $\BPROC$ and $\WPROC$ is not
  Archimedean.
\end{enumerate}
\end{exmp}

Notice that in the above example, it is essential that
$\starpref{\acta}\nil$ is not normed. Using that both norm and depth
are additive with respect to parallel composition, it follows that
$\ord$ is Archimedean on normed and bounded behaviour, and using that
a process expression with a norm or depth equal to $0$ is
behaviourally equivalent to $\nil$, it follows that $\nil$ is the
least element with respect to $\ord$ on normed and bounded behaviour.

\begin{prop} \label{prop:ordwn}
  The partial order $\ord$ on $\BPROCtn$, $\BPROCfin$, $\WPROCtn$ and
  $\WPROCfin$ is Archimedean and $\nil$ is its least element.
\end{prop}

\begin{exmp}\label{exa:notscompatible}
  Consider the process expressions
     $\cptermP=\starpref{\acta}\pref{\silent}\pref{\actb}\nil$ and
     $\cptermQ=\pref{\actb}\nil$ discussed in
   \Example{wninsufficient} (see also
   \Figure{counterexample}). Then,  since
  $\cptermP\step{\silent}\cptermQ$ and
  $\eqclass{\cptermP}\neq\eqclass{\cptermQ}$, we have that
  $\eqclass{\cptermQ}\sord\eqclass{\cptermP}$, but also
  $\eqclass{\cptermQ}\merge\eqclass{\cptermP}
   =\eqclass{\cptermP}\merge\eqclass{\cptermP}$.
  It follows that $\ord$ is not strictly compatible in $\BPROC$,
  $\BPROCtn$, $\WPROC$, and $\WPROCtn$.
\end{exmp}

We should now still ask ourselves the question whether $\ord$ on
$\BPROCfin$ and $\WPROCfin$ is strictly compatible. An important step
towards proving the property for, e.g., $\BPROCfin$ would be to
establish the following implication for all bounded process expressions $\cptermP$,
$\cptermQ$ and $\cptermR$:
\begin{equation*}
  \cptermP\step{\silent}\cptermQ\ \&\
  \cptermP\merge\cptermR\bbisim\cptermQ\merge\cptermR
\Longrightarrow
  \cptermP\bbisim\cptermQ
\enskip.
\end{equation*}
\Example{notscompatible}
illustrates that this implication does not hold for all normed
processes, suggesting that the implication is perhaps hard to
establish from first principles. In fact, all our attempts in this direction so
far have failed. Note, however, that establishing the implication
would be straightforward if we could use that $\merge$ is cancellative (i.e.,
$\cptermP\merge\cptermR\bbisim\cptermQ\merge\cptermR$ implies
$\cptermP\bbisim\cptermQ$), and this, in turn, would be easy if we
could use that $\BPROCfin$ has unique decomposition.

The difficulty of establishing strict compatibility is really with
strictness. Due to the shape of the operational rules for parallel
composition (see \Table{tss}) it is actually straightforward to
establish the following non-strict variant. Let $\Mon$ be a partial
commutative monoid; a partial order $\ord$ on $\Mon$ is
\emph{compatible} if for all $\monx,\mony,\monz\in\Mon$:
\begin{equation*}
  \text{if $\monx\ord\mony$ and $\mony\comp\monz$ is defined,
           then $\monx\comp \monz\ord\mony\comp\monz$.}
\end{equation*}

\begin{prop} \label{prop:compatible}
  The partial order $\ord$ on $\BPROCtn$, $\BPROCfin$,
  $\WPROCtn$, and $\WPROCfin$ is \emph{compatible}.
\end{prop}

A partial order on a partial commutative monoid that has all the
properties of a decomposition order except that it is compatible
instead of strictly compatible, we shall henceforth call a \emph{weak}
decomposition order.
\begin{defn} \label{def:wdecomporder}
  Let $\Mon$ be a partial commutative monoid; a partial order $\ord$
  on $\Mon$ is a \emph{weak decomposition order} if it is
  well-founded, has the identity element $\emcomp\in\Mon$ as least
  element, is compatible, precompositional and Archimedean.
\end{defn}

The following corollary summarises
Propositions~\ref{prop:ordgeneral}, \ref{prop:ordwn} and \ref{prop:compatible}.
\begin{cor} \label{cor:wdecompord}
  The partial order $\ord$ on $\BPROCtn$, $\BPROCfin$, $\WPROCtn$,
  and $\WPROCfin$ is a weak decomposition order.
\end{cor}

In \cite{LO05} it is proved that the existence of a decomposition
order is a sufficient condition for a partial commutative monoid to
have unique decomposition. Note that, since $\ord$ is a weak
decomposition order on $\BPROCtn$ and $\WPROCtn$, and since according
to \Example{wninsufficient} these commutative monoids do not have
unique decomposition, the existence of a \emph{weak} decomposition order is
\emph{not} a sufficient condition for having unique decomposition; it
should be supplemented with additional requirements to get a
sufficient condition.

Strictness of compatibility---which is the only
difference between the notion of decomposition order of \cite{LO05}
and the notion of weak decomposition order put forward here---is used
in \cite{LO05} both in the proof of \emph{existence} of decompositions
and in the proof that decompositions are  \emph{unique}. Thanks to the
strengthening of the Archimedean property (cf.\ \Remark{archimedean}),
it is possible to establish the existence of decompositions in partial
commutative monoids endowed with a weak decomposition order.

\begin{prop} \label{prop:wdecompexistence}
  In every partial commutative monoid with a weak decomposition order,
  every element of $\Mon$ has a decomposition.
\end{prop}
\begin{proof}
  Let $\Mon$ be a commutative monoid with a weak decomposition order
  $\ord$; we prove with $\ord$-induction that every element $\Mon$ has
  a decomposition.
  Let $\monx$ be an element of $\Mon$ and suppose, by way of induction
  hypothesis, that all $\ord$-predecessors of $\monx$ have a
  decomposition; we distinguish two cases:
  \begin{enumerate}
  \item Suppose there exist $\mony,\monz\sord\monx$ such that
    $\monx=\mony\comp\monz$. Then by the induction hypothesis $\mony$
    and $\monz$ have decompositions $\dcd[\mony]$ and $\dcd[\monz]$,
    respectively, and their sum $\dcd[\mony]\multiplus\dcd[\monz]$
    (see \Example{pcm} for the definition of
    $\multiplus$) is a decomposition of $\monx$. \item Suppose there
    do not exist $\mony,\monz\sord\monx$ such that
    $\monx=\mony\comp\monz$. Then, for all $\mony,\monz\in\Mon$ such
    that $\monx=\mony\comp\monz$ we have that $\mony=\monx$ or
    $\monz=\monx$, and hence $\monx=\mony\comp\monx$ or
    $\monx=\monx\comp\monz$.

    On the one hand, from $\monx=\mony\comp\monx$ it follows
    that $\monx=\mony^n\comp\monx$ for all $n\in\N$.
    Hence, since $\emcomp$ is the least element of $\Mon$ with respect
    to $\ord$ and $\ord$ is compatible, we have
      $\mony^n = \emcomp\comp\mony^n \ord \monx\comp\mony^n = \mony^n\monx=\monx$
    for all $n\in\N$.
    So by the Archimedean property it follows that $\mony=\emcomp$.
    On the other hand, from $\monx=\monx\comp\monz$ it follows by a
    similar argument that $\monz=\emcomp$.

    Thus, we have now established that $\monx=\mony\comp\monz$ implies
    $\mony=\emcomp$ or $\monz=\emcomp$ for all $\mony,\monz\in\Mon$.
    It follows that either $\monx=\emcomp$, in which case it  has the
    empty multiset $\emptymset$ as decomposition, or $\monx$ is
    indecomposable, in which case it has $\makemset{\monx}$ as
    decomposition.
\qedhere
\end{enumerate}
\end{proof}

It follows from \Corollary{wdecompord} and
\Proposition{wdecompexistence} that in the monoids $\BPROCtn$,
$\BPROCfin$, $\WPROCtn$, and $\WPROCfin$ every element has a
decomposition. In the next section, we shall propose a
general subsidiary property that will allow us to establish uniqueness
of decompositions in commutative monoids with a weak decomposition
order; in \Section{application} we shall establish that this property
holds in $\BPROCfin$ and $\WPROCfin$ and conclude that these monoids
have unique decomposition.

\section{Uniqueness} \label{sec:uniqueness}

The failure of $\ord$ on $\BPROCfin$ and $\WPROCfin$ to be strictly
compatible prevents us from getting our unique decomposition results
for those commutative monoids as an immediate consequence
of the result in \cite{LO05}. Nevertheless, many of the ideas in the
proof of uniqueness of decompositions in \cite{LO05} can be adapted
and reused in the context of commutative monoids endowed with a weak
decomposition order. Most importantly, the crucial \emph{subtraction}
property of decomposition orders holds for weak decomposition orders
too, for its proof (see the proofs of Lemmas 24 and 25 and Corollary 16 in
\cite{LO05}) does not rely on strictness of compatibility.%
\footnote{The proof of Lemma 24 in \cite{LO05} does refer to Proposition 23 of
  \cite{LO05} in which it is established that in a partial commutative monoid with a
  decomposition order every element has a decomposition using strict
  compatibility. But we have established in
  \Proposition{wdecompexistence} that strictness of compatibility is
  not needed to conclude that every element has a decomposition.}

\begin{lem}[Subtraction] \label{lem:subtraction}
  Let $\Mon$ be a partial commutative monoid with a weak decomposition
  order $\ord$.
  Let $\monx,\mony,\monz\in\Mon$, and suppose that $\monx\comp\mony$
  has a unique  decomposition.
  Then
    $\monx\comp\mony\sord\monx\comp\monz$
  implies
    $\mony\sord\monz$.
\end{lem}

This section is devoted to eliminating the use of strictness of
compatibility from most of the argument in \cite{LO05} showing partial
commutative monoids endowed with a decomposition order have unique
decomposition, at the expense of more involved technical details.Thus,
we shall push the use of strictness of compatibility to one corner
case that can be settled if we replace strictness of compatibility by
an alternative requirement referred to as \emph{power
  cancellation}. This will culminate in \Theorem{uniqueness}, the main
result of this article, which states that a partial commutative monoid
endowed with a weak decomposition order satisfying power cancellation
has unique decomposition. In \Section{application} we shall then
establish that the weak decomposition orders that we have already
defined on $\BPROCfin$ and $\WPROCfin$ satisfy power cancellation and
conclude that these monoids both have unique decomposition.

Let us fix, for the remainder of this section, a partial commutative
monoid $\Mon$ and a weak decomposition order $\ord$ on $\Mon$.

The uniqueness proof in \cite{LO05} considers a minimal counterexample
against unique decomposition, i.e., an element of the commutative
monoid with at least two distinct decompositions, say $\dcd[1]$ and
$\dcd[2]$, that is $\ord$-minimal in the set of all such
elements. Then, an important technique in the proof is to select a
particular indecomposable in one of the two decompositions and replace
it by predecessors with respect to the decomposition order. From
minimality together with strict compatibility it is then concluded
that the resulting decomposition is unique, which plays a crucial role
in subsequent arguments towards a contradiction.  To avoid the use of
strictness of compatibility, we need a more sophisticated notion of
minimality for the considered counterexample. The idea is to not just
pick a $\ord$-minimal element among the elements with two or more
decompositions; we also choose the presupposed pair of distinct
decompositions $(\dcd[1],\dcd[2])$ in such a way that it is minimal
with respect to a well-founded ordering induced by $\ord$ on them.

\paragraph{The decomposition extension of $\ord$}

Let $X$ be a set.
In \Example{pcm} we introduced the notion of
  multiset over $X$ together with the binary operation $\multiplus$
  for multiset sum; we also
  need multiset difference: 
If $\msetm$ and $\msetn$ are multisets over $X$, then
their multiset difference is the multiset
$\msetm\multimin\msetn$ that satisfies, for all $x\in X$,
\begin{equation*}
  \msetm\multimin\msetn(x) =
    \left\{\begin{array}{ll}
      \msetm(x)-\msetn(x)
        & \quad\text{if $\msetm(x)\geq\msetn(x)$\enskip;}\\
      0 & \quad\text{otherwise}\enskip.
    \end{array}\right.
\end{equation*}
We define the \emph{decomposition extension} $\sdord$ of $\sord$
by $\dcd\sdord\dcd'$ if, and only if, there exist, for some $k\geq 1$, a
sequence of indecomposables $\procp[1],\dots,\procp[k]\in\Mon$, a
sequence $\monx[1],\dots,\monx[k]\in\Mon$, and a sequence of
decompositions $\dcd[1],\dots,\dcd[k]$ such that
\begin{enumerate}
\renewcommand{\theenumi}{\roman{enumi}}
\renewcommand{\labelenumi}{(\theenumi)}
\item $\monx[i]\sord\procp[i]$ ($1\leq i \leq k$);
\item each $\dcd[i]$ is a decomposition of $\monx[i]$ ($1\leq i
  \leq k$); and
\item $\dcd=(\dcd'-\makemset{\procp[1],\dots,\procp[k]})
              \multiplus(\dcd[1]\multiplus\cdots\multiplus\dcd[k]$).
\end{enumerate}
We write $\dcd\dord\dc[d']$ if $\dcd=\dc[d']$ or
$\dcd\sdord\dc[d']$. Note that if $\dcd\dord\dc[d']$, $\monx$ is
the composition of $\dcd$, and $\mony$ is the composition of
$\dc[d']$, then, by compatibility, $\monx\ord\mony$.

The following two lemmas express general properties of the
decomposition extension.

\begin{lem} \label{lem:decsimulation}
  Let $\dcd[1]$ and $\dcd[2]$ be decompositions such that $\dcd[1]\dceq\dcd[2]$.
  Then for every decomposition $\dc[d_1']\dord\dcd[1]$ there exists a
  decomposition $\dc[d_2']\dord\dcd[2]$ such that
  $\dc[d_1']\dceq\dc[d_2']$.
\end{lem}
\begin{proof}
  Let $\dc[d_1']\dord\dcd[1]$.
  Clearly, if $\dc[d_1']=\dcd[1]$, then we can take $\dc[d_2']=\dcd[2]$
  and immediately get $\dc[d_1']=\dcd[1]\dceq\dcd[2]=\dc[d_2']$.  So
  suppose that $\dc[d_1']\sdord\dcd[1]$.  Then there exist a
  decomposition $\dc[d_1'']$ and, for some $k\geq 1$, sequences of
  indecomposables $\primp_1,\dots,\primp_k$ and of decompositions
  $\dcd[1,1],\dots,\dcd[1,k]$ such that
  \begin{enumerate}
  \renewcommand{\theenumi}{\roman{enumi}}
  \renewcommand{\labelenumi}{(\theenumi)}
  \item $\dcd[1]=\dc[d_1'']\multiplus\makemset{\primp_1,\dots,\primp_k}$;
  \item each $\dcd[1,i]$ is a decomposition of a predecessor of
    $\primp_i$ ($1\leq i \leq k$); and
  \item $\dc[d_1']=\dc[d_1'']\multiplus\dcd[1,1]\multiplus\cdots\multiplus\dcd[1,k]$.
  \end{enumerate}
  Denote by
    $\monx$, $\monx'$, $\monx''$, and $\monx[i]$ ($1\leq i \leq k$)
  the compositions of the decompositions
     $\dcd[1]$, $\dc[d_1']$, $\dc[d_1'']$  and $\dcd[1,i]$ ($1\leq i \leq k$),
  respectively.
  Then, for all $1\leq i \leq k$, $\monx[i]$ is a $\ord$-predecessor
  of $\primp_i$, so, by compatibility,
  \begin{equation*}
    \monx'=\monx''\comp\monx[1]\comp\cdots\comp\monx[k]
  \ord
    \monx''\comp\primp_1\comp\cdots\comp\primp_k=\monx
  \enskip.
  \end{equation*}
  If $\monx'=\monx$, then $\dc[d_1']\dceq\dcd[2]$, so we can take
  $\dc[d_2']=\dcd[2]$ and get $\dc[d_1']\dceq\dc[d_2']$.

  It remains to consider the case that $\monx'\sord\monx$.
  Let $\primq_1,\dots,\primq_{\ell}$ be such that
    $\dcd[2]=\makemset{\primq_1,\dots,\primq_{\ell}}$.
  Then, since $\monx$ is the composition of $\dcd[1]$ and
  $\dcd[1]\dceq\dcd[2]$, 
     $\monx=\primq_1\comp\cdots\comp\primq_\ell$, and hence, by
  precompositionality, there exist $\mon[x_1'],\dots,\mon[x_{\ell}']$
  such that $\monx'=\mon[x_1']\comp\cdots\comp\mon[x_{\ell}']$ and
  $\mon[x_i']\ord\primq_i$ ($1\leq i \leq \ell$).  Note that, since
  $\monx'\neq\monx$, there is at least one $1\leq i \leq \ell$ such
  that $\monx[i]'\neq\primq_i$. 
  We assume without loss of generality that the indecomposables
    $\primq_1,\dots,\primq_{\ell}$ 
  and their weak predecessors $\mon[x_1'],\dots,\mon[x_{\ell}']$ are
  ordered in such a way that there exists $1 \leq j \leq \ell$ such that
  $\mon[x_i']=\primq_i$ for all $1 \leq i < j$, and
  $\monx[i]'\sord\primq_i$ for all $j \leq i \leq \ell$.  Let
  $\dc[d_{2,j}'],\dots,\dc[d_{2,\ell}']$ be decompositions of
  $\mon[x_j'],\dots,\mon[x_{\ell}']$, and define
  $\dc[d_2']=\makemset{\primq_{1},\dots,\primq_{j-1}}
  \multiplus\dc[d_{2,j}']\multiplus\cdots\multiplus\dc[d_{2,\ell}']$.
  Then $\dc[d_2']\sdord\dcd[2]$, and since $\dc[d_1']$ and
  $\dc[d_2']$ both have $\monx'$ as their composition, we have that
  $\dc[d_1']\dceq\dc[d_2']$.
\end{proof}

\begin{lem}
  The relation $\dord$ is a well-founded partial order on decompositions.
\end{lem}
\begin{proof}
  It is immediate from the definition of $\dord$ that it is reflexive
  and transitive. It remains to establish that $\dord$ is
  well-founded, for a well-founded reflexive and transitive relation
  is a partial order. To this end, note that $\dord$ is a subset of
  the standard multiset ordering associated with the well-founded
  partial order $\ord$, which is proved to be well-founded by
  Dershowitz and Manna in \cite{DM79}.
\end{proof}

We shall use the well-foundedness of both $\ord$ and the Cartesian
order $\Cdordsym$ induced on pairs of decompositions by the
well-founded partial order $\dord$.  For
two pairs of decompositions $(\dcd[1],\dcd[2])$ and $(\dc[d_1'],\dc[d_2'])$, 
we write $(\dcd[1],\dcd[2])\Cdord(\dc[d_1'],\dc[d_2'])$ if
$\dcd[1]\dord\dc[d_1']$ and $\dcd[2]\dord\dc[d_2']$.  A pair of
decompositions $(\dcd[1],\dcd[2])$ is said to be a
\emph{counterexample} against unique decomposition if $\dcd[1]$ and
$\dcd[2]$ are distinct but equivalent, i.e., if $\dcd[1]\dceq\dcd[2]$,
but not $\dcd[1]=\dcd[2]$. A counterexample $(\dcd[1],\dcd[2])$
against unique decomposition is \emph{minimal} if it is both minimal
with respect to $\ord$ and minimal with respect to $\Cdord$.
That is, a counterexample $(\dcd[1],\dcd[2])$ against unique
decomposition is \emph{minimal} if
\begin{enumerate}
\item all $\ord$-predecessors of the (common) composition of $\dcd[1]$
  and $\dcd[2]$ have a unique decomposition; and
\item for all $(\dc[d_1'],\dc[d_2'])$ such that
      $(\dc[d_1'],\dc[d_2'])\Cdord(\dcd[1],\dcd[2])$ and 
    $(\dc[d_1'],\dc[d_2'])\neq(\dcd[1],\dcd[2])$ it holds that
    $\dc[d_1']\dceq\dc[d_2']$ implies $\dc[d_1']=\dc[d_2']$.
\end{enumerate}

Since both $\ord$ and $\Cdordsym$ are well-founded, if unique
decomposition would fail, then there would exist a minimial
counterexample. The general idea of the proof is that we derive a
contradiction from the assumption that there exists a minimal
counterexample $(\dcd[1],\dcd[2])$ against unique decomposition.  The
decompositions $\dcd[1]$ and $\dcd[2]$ should be distinct, so the set
of indecomposables that occur more often in one of the decompositions
than in the other is non-empty. This set is clearly also finite, so it
has $\ord$-maximal elements. We declare $\primp$ to be such a
$\ord$-maximal element, and assume, without loss of generality, that
$\primp$ occurs more often in $\dcd[1]$ than in $\dcd[2]$. Then we
have that
\begin{enumerate}
\renewcommand{\theenumi}{\Alph{enumi}}
\renewcommand{\labelenumi}{(\theenumi)}
  \item \label{ass:A}
    $\dcd[1](\primp)>\dcd[2](\primp)$; and
  \item \label{ass:B}
    $\dcd[1](\primq)=\dcd[2](\primq)$ for all indecomposables
    $\primq$ such that $\primp\sord\primq$.
\end{enumerate}
We shall distinguish two cases, based on how the difference between
$\dcd[1]$ and $\dcd[2]$ manifests itself, and derive a contradiction
in both cases:
\begin{enumerate}
\item $\dcd[1](\primp)>\dcd[2](\primp)+1$ or $\dcd[1](\primq)\neq 0$
  for some indecomposable $\primq$ distinct from $\primp$; we refer to
  this case by saying that $\dcd[1]$ and $\dcd[2]$ are \emph{too far
    apart}.
\item $\dcd[1](\primp)=\dcd[2](\primp)+1$ and $\dcd[1](\primq)=0$ for
  all $\primq$ distinct from $\primp$; we refer to this case by saying
  that $\dcd[1]$ and $\dcd[2]$ are \emph{too close together}.
\end{enumerate}

\paragraph{Case 1: $\dcd[1]$ and $\dcd[2]$ are too far apart}
In this case, either the multiplicity of $\primp$ in $\dcd[1]$ exceeds
the multiplicity of $\primp$ in $\dcd[2]$ by at least $2$, or the
difference in multiplicities is $1$ but there is another
indecomposable $\primq$, distinct from $\primp$, in $\dcd[1]$.
We argue that $\dcd[1]$ has a predecessor $\dc[d']$ in which
$\primp$ occurs more often than in any predecessor of
$\dcd[2]$, while, on the other hand, the choice of a minimal
counterexample implies that every predecessor of $\dcd[1]$ is
also a predecessor of $\dcd[2]$. (The arguments leading to a
contradiction in this case are analogous to the arguments in the proof
in \cite{LO05}; the only important difference is the use of the ordering
$\dord$ instead of $\ord$.)

Using that $\ord$ is compatible and Archimedean, it can be established
that there is a bound on the multiplicity of $\primp$ in the
predecessors of $\dcd[1]$.
\begin{lem}
  The set $\{\dc[d_1'](\primp):\dc[d_1']\sdord\dcd[1]\}$
  is finite.
\end{lem}
\begin{proof}
  Denote by $\mony$ the composition of $\dcd[1]$.
  Clearly, if $\dc[d_1']$ is a predecessor of $\dcd[1]$, then, by
  compatibility, we have that
     $\primp^n\ord\mony$ for all $n\leq \dc[d_1'](\primp)$.
  Hence, if the set
    $\{\dc[d_1'](\primp):\dc[d_1']\sdord\dcd[1]\}$
  would \emph{not} be finite, then $\primp^n\ord\mony$ for all
  $n\in\N$, from which, since $\ord$ is Archimedean, it would follow
  that $\primp=\emcomp$. But $\primp=\emcomp$ is in contradiction with
  the assumption that $\primp$ is indecomposable.
\end{proof}

Let $m$ be the maximum of the multiplicities of $\primp$ in
predecessors of $\dcd[1]$, i.e.,
\begin{equation} \label{eq:mdef}
  m := \max \{\dc[d_1'](\primp):\dc[d_1']\sdord\dcd[1]\}
\enskip.
\end{equation}
On the one hand, if $\dcd[1](\primp)>\dcd[2](\primp)+1$, then
  $\dcd[1]-\makemset{\primp}\sdord\dcd[1]$,
so $m\geq \dcd[1](\primp)-1 >\dcd[2](\primp)$.
On the other hand, if $\dcd[1](\primq)\neq 0$ for some indecomposable
$\primq\neq\primp$, then $\dcd[1]-\makemset{\primq}\sdord\dcd[1]$, so
$m\geq\dcd[1](\primp)>\dcd[2](\primp)$.
Hence
\begin{equation} \label{eq:dcd2m}
  \dcd[2](\primp)<m
\enskip.
\end{equation}

If $k\in\N$, then we write $k\multimaal\makemset{\primp}$
for the decomposition consisting of $k$ occurrences of $\primp$, i.e.,
the multiset for which it holds that
\begin{equation*}
  (k \multimaal\makemset{\primp})(\primq)=
     \left\{\begin{array}{ll}
        k & \quad\text{if $\primq=\primp$\enskip;}\\
        0 & \quad\text{otherwise}\enskip.
    \end{array}\right.
\end{equation*}
From \eqref{eq:mdef} it is clear that
  $m\multimaal\makemset{\primp}\sdord\dcd[1]$.
Hence, by \Lemma{decsimulation}, there exists $\dc[d_2']\dord\dcd[2]$
such that $m\multimaal\makemset{\primp}\dceq\dc[d_2']$. Since
$(\dcd[1],\dcd[2])$ is a minimal counterexample, it follows that
$m\multimaal\makemset{\primp}=\dc[d_2']$, and hence
  $m\multimaal\makemset{\primp}\dord\dcd[2]$.
Moreover, from \eqref{eq:dcd2m}, it is clear that
$m\multimaal\makemset{\primp}\neq\dcd[2]$, so
$m\multimaal\makemset{\primp}\sdord\dcd[2]$.
Thus, we now have
\begin{equation} \label{eq:powerbelow}
  m\multimaal\makemset{\primp}
   \sdord
  \dcd[1],
  \dcd[2]
\enskip.
\end{equation}

We now proceed to argue that indecomposables distinct from
$\primp$ in $\dcd[2]$ can be used to \emph{create}
  $m-\dcd[2](\primp)$
additional occurrences of $\primp$ in predecessors of
$\dcd[2]$. Since those extra occurrences of $\primp$ can only be
created by occurrences in $\dcd[2]$ of indecomposables $\primq$ such
that $\primp\sord\primq$, it can be concluded by
assumption~(\ref{ass:B}) that $\dcd[1]$ must have the same potential for
creating extra occurrences of $\primp$. We shall see that this
reasoning will eventually lead to a contradiction with our  definition
of $m$ as the maximal number of occurrences of $\primp$ in
predecessors of $\dcd[1]$.

In the remainder of our argument, it will be convenient to have
notation for specific parts of $\dcd[1]$ and $\dcd[2]$: For $i=1,2$
we denote by $\dc[d_i^{\invsord\primp}]$ the multiset consisting of
all indecomposables $\primq$ in $\dcd[i]$ such that
$\primp\sord\primq$, i.e., $\dc[d_i^{\invsord\primp}]$ is defined by
\begin{equation*}
  \dc[d_i^{\invsord\primp}](\primq)
     =\left\{\begin{array}{ll}
          \dcd[i](\primq) & \quad\text{if $\primq\invsord\primp$}\enskip;\\
          0 & \quad\text{otherwise}\enskip;
       \end{array}\right.
\end{equation*}
we denote by $\dc[d_i^{=\primp}]$ the multiset consisting of all
occurrences of $\primp$ in $\dcd[i]$, i.e., $\dc[d_i^{=\primp}]$ is
defined by
\begin{equation*}
  \dc[d_i^{=\primp}](\primq)
     =\left\{\begin{array}{ll}
          \dcd[i](\primq) & \quad\text{if $\primq=\primp$}\enskip;\\
          0 & \quad\text{otherwise}\enskip;
       \end{array}\right.
\end{equation*}
we denote by $\dc[d_i^{\not\invord\primp}]$ the multiset consisting of all
occurrences of $\primp$ in $\dcd[i]$, i.e., $\dc[d_i^{\not\invord\primp}]$ is
defined by
\begin{equation*}
  \dc[d_i^{\not\invord\primp}](\primq)
     =\left\{\begin{array}{ll}
          \dcd[i](\primq) & \quad\text{if $\primq\not\invord\primp$}\enskip;\\
          0 & \quad\text{otherwise}\enskip.
       \end{array}\right.
\end{equation*}
Then clearly we have
\begin{equation*}
  \dcd[i]=\dc[d_i^{\invsord\primp}]\multiplus
  \dc[d_i^{=\primp}]\multiplus \dc[d_i^{\not\invord\primp}]
\qquad(i=1,2)
\enskip.
\end{equation*}

That the decompositions $\dc[d_i^{\invsord\primp}]$ ($i=1,2$)
both incorporate the potential of creating $m-\dcd[2](\primp)$ occurrences
of $\primp$ is formalised by proving that
  $(m-\dcd[2](\primp))\multimaal\makemset{\primp}
      \sdord\dc[d_i^{\invsord\primp}]$ ($i=1,2$).
We shall first prove
  $(m-\dcd[2](\primp))\multimaal\makemset{\primp}
      \sdord\dc[d_2^{\invsord\primp}]$,
and for this we need the following general lemma.

Let $\monx=\primp^{\dcd[2](\primp)}$ and
$\mony=\primp^{m-\dcd[2](\primp)}$, and denote by $\monz$ the
composition of
  $\dc[d_2^{\invsord\primp}]\multiplus \dc[d_2^{\not\invord\primp}]$.
Then $\monx\comp\mony=\primp^m$ and $\monx\comp\monz$ is equal to the 
composition of both $\dcd[1]$ and $\dcd[2]$. Note that from
\eqref{eq:powerbelow} and the minimality of the counterexample
$(\dcd[1],\dcd[2])$ it immediately follows that
$\primp^m=\monx\comp\mony\sord\monx\comp\monz$.
(For if $\monx\comp\mony=\monx\comp\monz$, then
$\primp^m=\monx\comp\monz$, so both
$(m\multimaal\makemset{\primp},\dcd[2])$ and
$(\dcd[1],m\multimaal\makemset{\primp})$ would constitute smaller
counterexamples.) It follows that $\monx\comp\mony$ has a unique
decomposition, so by \Lemma{subtraction} it follows that
  $\mony\sord\monz$.

\begin{lem}
  If $\mony \sord \monz$ and $\dcd$ is a decomposition of $\monz$,
  then $\mony$ has a
  decomposition $\dcd'$ such that $\dcd'\sdord\dcd$.
\end{lem}
\begin{proof}
  Let $\dcd=\makemset{\primp[1],\dots,\primp[k]}$.
  Then
    $\mony\sord \primp[1]\comp\cdots\comp\primp[k]$,
  so, by precompositionality, there exist
    $\mony[1],\dots,\mony[k]$
  such that
    $\mony[i]\ord\primp[i]$ for all $1\leq i \leq k$.
  The $\mony[i]$ have decompositions, say $\dcd[i]'$, and clearly
    $\dcd'=\dcd[1]'\multiplus\cdots\multiplus\dcd[k]'$
  is a decomposition of $\mony$ satisfying $\dcd'\dord\dcd$. Since
  $\dcd'=\dcd$ would imply $\mony=\monz$, it follows that
  $\dcd'\sdord\dcd$.
\end{proof}

By the preceding lemma, $\mony$ has a decomposition, say $\dc[d_2']$,
such that
  $\dc[d_2']\sdord
      \dc[d_2^{\invsord\primp}]\multiplus \dc[d_2^{\not\invord\primp}]$,
and since $\mony$, in fact, has a unique decomposition, it follows
that
  $(m-\dcd[2](\primp))\multimaal\makemset{\primp}
      \sdord
      \dc[d_2^{\invsord\primp}]\multiplus
      \dc[d_2^{\not\invord\primp}]$.
By definition of $\dc[d_2^{\not\invord\primp}]$, $\primp$ does not occur
in $\dc[d_2^{\not\invord\primp}]$, nor in any decomposition of a
predecessor of an indecomposable in $\dc[d_2^{\not\invord\primp}]$, so
  $(m-\dcd[2](\primp))\multimaal\makemset{\primp}
      \sdord
    \dc[d_2^{\invsord\primp}]$.
Since $\dc[d_2^{\invsord\primp}]=\dc[d_1^{\invsord\primp}]$ according
to assumption~(\ref{ass:B}), we have
  $(m-\dcd[2](\primp))\multimaal\makemset{\primp}
       \sdord\dc[d_1^{\invsord\primp}]$.
It follows that
\begin{equation} \label{eq:majorize}
   (m-\dcd[2](\primp)+\dcd[1](\primp))\multimaal\makemset{\primp}
      \sdord\dcd[1]
\enskip,
\end{equation}
and since
  $\dcd[2](\primp)<\dcd[1](\primp)$
according to assumption~(\ref{ass:A}), we find
  $m-\dcd[2](\primp)+\dcd[1](\primp)>m$.
Thus we have now derived a contradiction with our definition of $m$ in
\eqref{eq:mdef} as the maximum of the multiplicities of $\primp$ in
the predecessors of $\dcd[1]$.

\paragraph{Case 2: $\dcd[1]$ and $\dcd[2]$ are too close together}

In this case $\dcd[1]$ contains no other indecomposables
than $\primp$, while $\dcd[2]$ has $\dcd[1](\primp)-1$ occurrences of
$\primp$ supplemented with a multiset $\dc[d_2']$ of other
indecomposables.

In \cite{LO05} it is proved, via a sophisticated argument, that the
composition of $\dc[d_2']$ is a $\ord$-predecessor of $\primp$. Hence,
by strict compatibility, the composition of $\dcd[2]$ is an
$\ord$-predecessor of $\dcd[1]$, which is in contradiction with the
assumption that the decompositions $\dcd[1]$ and $\dcd[2]$ are
equivalent.

That $\ord$ is not strictly compatible, but just compatible, leaves
the possibility that $\dcd[1]$ and $\dcd[2]$ are equivalent even if
the composition of $\dc[d_2']$ is a  predecessor of
$\primp$. For $\BPROCfin$ and $\WPROCfin$ this possibility can be
ruled out by noting that the composition of $\dc[d_2']$ can be reached
from $\primp$ by $\silent$-transitions, and proving that every
transition of $\primp$ can be simulated by a transition of the
composition of $\dc[d_2']$. The following notion formalises this
reason in the abstract setting of commutative monoids with a weak
decomposition order.

\begin{defn}
  Let $\Mon$ be a partial commutative monoid, and let $\ord$ be a weak
  decomposition order on $\Mon$. We say that $\ord$ satisfies
  \emph{power cancellation} if for all $\monx,\mony\in\Mon$ and for every
  indecomposable  $\primp\in\Mon$ such that
  $\primp\not\sord\monx,\mony$ it holds that
  \begin{equation*}
    \primp^k\comp\monx=\primp^k\comp\mony
       \ \text{implies}\ \monx=\mony
       \ \text{for all $k\in\N$.}
  \end{equation*}
\end{defn}

Suppose that $\ord$ on $\Mon$ has power cancellation, let
$k=\dcd[2](\primp)$ and let $\monx$ be the composition of
$\dc[d_2']$. Then from $\dcd[1]\dceq\dcd[2]$ it follows that
\begin{equation*}
  \primp^k\comp\primp = \primp^k\comp\monx
\enskip.
\end{equation*}
Clearly, $\primp\not\sord\primp$ and, since $\dc[d_2']$ consists of
indecomposables $\primq$ such that $\primp\not\ord\primq$, it follows
by precompositionality that also $\primp\not\ord\monx$. Hence, since
$\ord$ has power cancellation, $\primp=\monx$, so $\dc[d_2']=\makemset{\primp}$. It
follows that $\dcd[1]=\dcd[2]$, which contradicts that
$(\dcd[1],\dcd[2])$ is a counterexample against unique decomposition.

\begin{thm} \label{thm:uniqueness}
  Every partial commutative monoid $\Mon$ with a weak decomposition order
  that satisfies power cancellation has unique decomposition.
\end{thm}

  We conclude this section with the observation that a weak
  decomposition satisfying power cancellation is, in fact, strictly
  compatible, and hence a decomposition order. To see this, consider a
  partial commutative monoid $\Mon$ endowed with a weak decomposition
  order $\ord$ that satisfies power cancellation. Since, by
  \Theorem{uniqueness}, $\Mon$ has unique decomposition, it also has
  \emph{cancellation}: if  $\monx\comp\monz=\mony\comp\monz$
  implies $\monx=\mony$ for all $\monx,\mony\in\Mon$ (see, e.g.,
  Corollary 19 in \cite{LO05}). Now, to establish that $\ord$ is
  strictly compatible, let $\monx, \mony, \monz \in \Mon$ and suppose
  that $\monx\sord\mony$ and $\mony\comp\monz$ is defined; then, by
  compatibility, $\monx\comp\monz\ord\mony\comp\monz$.
  Since $\monx\neq\mony$ implies $\monx\comp\monz\neq\mony\comp\monz$
  by cancellation, it follows that
    $\monx\comp\monz\sord\mony\comp\monz$.

\begin{cor}\label{cor:wdecomppcdecomp}
  A weak decomposition order satisfying power cancellation is a
  decomposition order.
\end{cor}

\section{Bounded behaviour has unique parallel
  decomposition} \label{sec:application}

In \Section{pcm} we have already established that in the
commutative monoids $\BPROCfin$ and $\WPROCfin$ every element has a
decomposition and that $\ord$ is a weak decomposition order on
$\BPROCfin$ and $\WPROCfin$. To be able to conclude from
\Theorem{uniqueness} at the end of \Section{uniqueness} that
$\BPROCfin$ and $\WPROCfin$ have unique 
decomposition, it remains to establish that $\ord$ on these commutative
monoids satisfies power cancellation.

\begin{prop}\label{prop:powercanc}
  The weak decomposition order $\ord$ on $\BPROCfin$ and $\WPROCfin$
  satisfies power cancellation.
\end{prop}
\begin{proof}
  We present the proof for $\BPROCfin$; the proof for $\WPROCfin$ is
  very similar except that some details are slightly simpler.

  Let $\primp$ be a indecomposable element in $\BPROCfin$, and let
  $\monx$, $\mony$ and $\monz$ be elements of $\BPROCfin$ such that
  $\primp\not\sord\monx,\mony$, and, for some $k\in\N$,
  \begin{equation} \label{eq:antecedent}
  \monz
  =
    \primp^k\comp\monx
  =
    \primp^k\comp\mony
  \enskip;
  \end{equation}
  we need to prove that $\monx=\mony$.

  To this end, we first note that the ordering $\ord$ on
    $\BPROCfin\times\BPROCfin\times\BPROCfin$
  defined by
  \begin{equation*}
    (u',v',w') \ord (u,v,w)\
      \text{if $u'\ord u$ and whenever $u'=u$ then also $v'\ord v$ and
        $w'\ord w$}
  \end{equation*}
  is well-founded.
  We proceed by $\ord$-induction on $(z,x,y)$, and suppose, by way of
  induction hypothesis, that whenever $(z',x',y')\sord (z,x,y)$ and, for
  some indecomposable element $\primp'\not\sord\monx',\mony'$ of $\BPROCfin$,
  $\monz'=(\primp')^{\ell}\comp \monx'=(\primp')^{\ell}\comp \mony'$, then $\monx'=\mony'$.

  Note that $\monx$ and $\mony$ are non-empty sets of process
  expressions, and that, to prove $\monx=\mony$, it suffices to show that
  there exist process expressions $\cptermQ\in\monx$ and
  $\cptermR\in\mony$ such that $\cptermQ\bbisim\cptermR$.
  By \Lemma{wellfounded}, the non-empty sets of process expressions
  $\monx$ and $\mony$ have minimal elements with respect to the
  ordering induced on process expressions by
  $(\step{\silent}^{*})^{-1}$.
  Let $\cptermQ$ and $\cptermR$ be $(\step{\silent}^{*})^{-1}$-minimal
  elements in $\monx$ and $\mony$, respectively;
  we prove that $\cptermQ\bbisim\cptermR$ by establishing that the
  binary relation
  \begin{equation*}
    \brelsym=\{(\cptermQ,\cptermR),(\cptermR,\cptermQ)\}\cup{\bbisim}
  \end{equation*}
  is a branching bisimulation.

  To this end, we first suppose that $\cptermQ\step{\act}\cptermQ'$ for some
  $\cptermQ'$, and prove that there exist $\cptermR''$ and $\cptermR'$
  such that 
    $\cptermR\ssteps{}\cptermR''\optstep{\act}\cptermR'$,
    $\cptermQ\brel\cptermR''$, and
    $\cptermQ'\brel\cptermR'$.

  Let $\cptermP$ be an element of $\primp$, denote by $\cptermP^{\underline{k}}$
  the $k$-fold parallel composition of $\cptermP$, and let
    $\monz'=\beqclass{\cptermP^{\underline{k}}\merge\cptermQ'}$.
  Then $\monz'\ord\monz$, so we can distinguish two cases:
  \begin{distinction}
  \Case{1} Suppose that $\monz'=\monz$.
     Then, since $\cptermP^{\underline{k}}\merge\cptermQ$ is bounded, it
     follows that $\act=\silent$. 
     Let $\monx'=\beqclass{\cptermQ'}$; since $\cptermQ$ is a minimal
     element of $\monx$, we have that $\monx'\sord\monx$.
     Hence, $(\monz',\monx',\mony)\sord(\monz,\monx,\mony)$,
     so by the induction hypothesis
       $\beqclass{\cptermQ'}=\monx'=\mony=\beqclass{\cptermR}$.
     It follows that  $\cptermQ'\bbisim\cptermR$, and we can take
     $\cptermR''=\cptermR'=\cptermR$.
  \Case{2} Suppose that $\monz'\sord\monz$. Then, by the induction
    hypothesis, $\ord$ on the partial commutative submonoid
    $\{\monz'':\monz''\ord\monz'\}$ of $\BPROCfin$ satisfies power
    cancellation. By \Theorem{uniqueness}, it follows that $\monz'$
    has a unique decomposition in that submonoid, and hence in
    $\BPROCfin$ too.
    From $\cptermQ\step{\act}\cptermQ'$ it follows that
    \begin{equation*}
      \cptermP^{\underline{k}}\merge\cptermQ\step{\act}\cptermP^{\underline{k}}\merge\cptermQ'
    \enskip,
    \end{equation*}
    and hence, since
      $\cptermP^{\underline{k}}\merge\cptermQ\bbisim\cptermP^{\underline{k}}\merge\cptermR$
    according to \eqref{eq:antecedent},
    there exist $\cptermR'$, $\cptermR''$, $\cpterm[S]'$, and
    $\cpterm[S]''$ such that
    \begin{equation*}
      \cptermP^{\underline{k}}\merge\cptermR
          \ssteps{}
        \cpterm[S]''\merge\cptermR''
          \optstep{\act}
         \cpterm[S]'\merge\cptermR'
    \enskip,
    \end{equation*}
    with
       $\cptermP^{\underline{k}}\merge\cptermQ\bbisim\cpterm[S]''\merge\cptermR''$
    and
       $\cptermP^{\underline{k}}\merge\cptermQ'\bbisim\cpterm[S]'\merge\cptermR'$.
    We have that
    \begin{equation*}
      \beqclass{\cptermR'}\ord\beqclass{\cptermR''}\ord\beqclass{\cptermR}
    \end{equation*}
    and
    \begin{equation*}
       \beqclass{\cpterm[S]'}\ord\beqclass{\cpterm[S]''}\ord\beqclass{\cptermP^k}
    \enskip,
    \end{equation*}
    and, since
      $\beqclass{\cpterm[S]'}\merge\beqclass{\cptermR'}
        =\monz'
       \neq\monz
       =\beqclass{\cptermP^{\underline{k}}\merge\cptermR}$,
    it also holds that
      $\beqclass{\cptermR'}\neq\beqclass{\cptermR}$,
    or
      $\beqclass{\cpterm[S]'}\neq\beqclass{\cptermP^{\underline{k}}}$.
     We distinguish two subcases:
     \begin{distinction}
     \Case{2.1}
        Suppose $\beqclass{\cptermR'}\sord\beqclass{\cptermR}$.
        Then, since $\primp\not\sord\monx=\beqclass{\cptermR}$, the
        unique decomposition of $\beqclass{\cptermR'}$ cannot
        have occurrences of $\primp$. Since $\monz'$ has $k$
        occurrences of $\primp$, it follows that
          $\primp^k
              \ord\beqclass{\cpterm[S]'}
              \ord\beqclass{\cpterm[S]''}
              \ord\beqclass{\cptermP^{\underline{k}}}
              =\primp^k$,
        so
          $\beqclass{\cpterm[S]'}
              =\beqclass{\cpterm[S]''}
              =\primp^k$.
        Since
          $\monz'=\primp^k\merge\beqclass{\cptermQ'}
                       =\primp^k\merge\beqclass{\cptermR'}$,
        by the induction hypothesis
          $\beqclass{\cptermQ'}=\beqclass{\cptermR'}$,
        and hence $\cptermQ'\brel\cptermR'$.

        It remains to establish that $\cptermQ\brel\cptermR''$.
        If $\cptermR''=\cptermR$, then, since $\cptermQ\brel\cptermR$,
        this is immediate.
        If $\cptermR''\neq\cptermR$, then since $\cptermR$ is a
        $(\step{\silent}^{*})^{-1}$-minimal element of $\mony$, it
        follows that
          $\beqclass{\cptermR''}\sord\beqclass{\cptermR}$,
        so from
          $\monz=\primp^k\merge\beqclass{\cptermQ}
                       =\primp^k\merge\beqclass{\cptermR''}$
        it follows by the induction hypothesis that
          $\beqclass{\cptermQ}=\beqclass{\cptermR''}$,
        and hence $\cptermQ\brel\cptermR''$.
     \Case{2.2}
        Suppose $\beqclass{\cpterm[S]'}\sord\beqclass{\cptermP^k}$.
        Then the multiplicity of $\primp$ in the unique decomposition
        of $\beqclass{\cpterm[S']}$ is at most $k-1$. Hence, since
          $\beqclass{\cpterm[S']}\merge\beqclass{\cpterm[R']}
              = z'
              = \primp^k\merge\beqclass{\cpterm[Q']}$,
        it follows that
        $\primp$ must be an element of $\beqclass{\cptermR'}$. This
        means that $\primp\ord\beqclass{\cptermR'}$, and since
        $\primp\not\sord\mony=\beqclass{\cptermR}$, 
        it follows that
          $\beqclass{\cptermP}=\primp=\beqclass{\cptermR}$,
        and hence
          $\cptermP \bbisim\cptermR$.
        Thus, we also get that the multiplicity of $\primp$ in the
        decomposition of $\beqclass{\cpterm[S]'}$ is, in fact, $k-1$,
        and therefore we can assume without loss of generality that
        there exist process expressions
        $\cptermP[1],\cptermP[2],\dots,\cptermP[k],\cptermP[1]'$ such
        that
        \begin{gather*}
            \cpterm[S]''=\cptermP[1]\merge\cptermP[2]\merge\cdots\merge\cptermP[k]
        \enskip,\\
            \cpterm[S]'=\cptermP[1]'\merge\cptermP[2]\merge\cdots\merge\cptermP[k]
        \enskip,\\
            \cptermP\ssteps\cptermP[i]\quad(1\leq i \leq k)\enskip,\\
            \cptermP\bbisim\cptermP[i]\quad(2\leq i\leq k)\enskip,\
         \text{and}\\
            \cptermP[1]\optstep{\act}\cptermP[1]'
         \enskip.
         \end{gather*}
         From
           $\cptermP^{\underline{k}}\merge\cptermQ'
                \bbisim
             \cptermR\merge\cptermP[2]\merge\cdots\cptermP[k]\merge\cptermP[1]'$,
            $\cptermP\bbisim\cptermR$, and
            $\cptermP\bbisim\cptermP[i]$ ($2\leq i \leq k$)
         it follows that $\cptermQ'\bbisim\cptermP[1]'$.
         Hence, since $\cptermP\bbisim\cptermR$, there exist
         $\cptermR[1]$, $\cptermR[1]'$ such that
           $\cptermR\ssteps\cptermR[1]\optstep{\act}\cptermR[1]'$,
           $\cptermP[1]\bbisim\cptermR[1]$,
         and
           $\cptermP[1]'\bbisim\cptermR[1]'$.
         From $\cptermQ'\bbisim\cptermP[1]'\bbisim\cptermR[1]'$ it
         follows that $\cptermQ'\brel\cptermR[1]'$.

         It remains to establish that $\cptermQ\brel\cptermR[1]$.
         If $\cptermR[1]=\cptermR$, then, since
         $\cptermQ\brel\cptermR$, this is immediate.
         If $\cptermR[1]\neq\cptermR$, then, since $\cptermR$ is a
         $(\step{\silent}^{*})^{-1}$-minimal element of $\mony$, it
         follows that
            $\beqclass{\cptermP[1]}
               =\beqclass{\cptermR[1]}
               \sord\beqclass{\cptermR}
               =\beqclass{\cptermP}$.
         So from
             $\monz
                 =\primp^k\merge\beqclass{\cptermQ}
                 =\primp^k\merge\beqclass{\cptermP[1]}$
         it follows by the induction hypothesis that
           $\cptermQ\bbisim\cptermP[1]\bbisim\cptermR[1]$,
         and hence $\cptermQ\brel\cptermR[1]$.
     \end{distinction}
  \end{distinction}
  In a completely analogous manner, it can be established that whenever
  $\cptermR\step{\act}\cptermR'$ for some process expression
  $\cptermR'$, then there exist process expressions $\cptermQ'$ and
  $\cptermQ''$ such that
     $\cptermQ\ssteps\cptermQ''\optstep{\act}\cptermQ'$,
     $\cptermR\brel\cptermQ''$, and
     $\cptermR'\brel\cptermQ'$.

  We conclude that $\brelsym$ is a branching bisimulation, and hence
    $\cptermQ\bbisim\cptermR$.
\end{proof}

By Corollaries~\ref{cor:wdecompord} and \ref{prop:powercanc}, the
commutative monoids $\BPROCfin$ and $\WPROCfin$ are endowed with a
weak decomposition order $\ord$ satisfying power cancellation. By
\Theorem{uniqueness} it follows that they have unique decomposition.

\begin{cor}
  The commutative monoids $\BPROCfin$ and $\WPROCfin$ have unique
  decomposition.
\end{cor}

\section{Concluding remarks} \label{sec:conclusions}

We have presented a general sufficient condition on partial
commutative monoids that implies the property of unique decomposition,
and is applicable to commutative monoids of behaviour incorporating a
notion of unobservability. We have illustrated the application of our
condition in the context of a very simple process calculus with an
operation for pure interleaving as parallel compostion. The
applicability is, however, not restricted to settings with this
particular type of parallel composition. In fact, it is to be expected
that our condition, similarly as in \cite{LO05}, can also be used to
prove unique decomposition results in settings with more complicated
notions of parallel composition operator allowing, e.g.,
synchronisation between components.

We leave for future investigations to what extent our theory of unique
decomposition can be applied to variants of $\pi$-calculus. The
article \cite{DELL13}, in which unique parallel decomposition is
established for a fragment of Applied $\pi$-calculus, will serve as a
good starting point. A complication, illustrated in \cite{DELL13}, is
that parallel components may fuse into a single indecomposable process
due to scope extrusion. As a consequence, precompositionality fails
for the order induced on equivalence classes of $\pi$-terms by
the transition relation. A solution may be to use a fragment of the
transition relation that avoids scope extrusion.

In \cite{BH10}, Balabonski and Haucourt address the problem of unique
parallel decomposition in the context of a concurrent programming
language with a geometric semantics. It is less clear whether our
general theory of unique decomposition is applicable there too; at
least, the geometric semantics does not as naturally induce a candidate
decomposition order on processes as in a process calculus with a
transition system semantics. It would be interesting to compare the
approaches. 

\paragraph{Acknowledgement} The author is grateful to both
  reviewers for their fine suggestions. In particular,
  one of the reviewers suggested \Proposition{wdecompexistence} and
  its proof, which resulted in a stronger formulation of the main
  result and significant improvements in presentation.

\end{document}